\documentclass[bimj,fleqn]{w-art}
\usepackage{times}
\usepackage{w-thm}
\usepackage{natbib}
\usepackage{mathrsfs}
\usepackage{booktabs}
\theoremstyle{plain}

\theoremstyle{definition}

\usepackage[]{graphicx}
\chardef\bslash=`\\ 
\usepackage{moreverb,url}
\usepackage{bm} 
\usepackage{bbm} 
\usepackage{soul}
\usepackage{mathrsfs}
\usepackage{enumitem} 
\usepackage{varwidth} 
\usepackage{booktabs}
\usepackage{multirow}

\usepackage{hyperref}
\hypersetup{colorlinks,bookmarksopen,bookmarksnumbered,citecolor=red,urlcolor=red}

\DOIsuffix{bimj.200100000}
\Volume{52}
\Issue{61}
\Year{2023}
\pagespan{1}{}

\newcommand{\nc}{\newcommand}
\newcommand{\lsq}{\left[}
\newcommand{\rsq}{\right]}
\newcommand{\lbc}{\left \{ }
\newcommand{\rbc}{\right \} }
\newcommand{\lp}{\left(}
\newcommand{\rp}{\right)}

\nc{\cond}{{\, \vert \,}}
\nc{\indep}{{\, \perp \! \! \! \perp  \,} }
\nc{\tsps}{^{ {\rm T} } }
\nc{\code}[1]{\texttt{#1}}
\nc{\R}{{\normalfont\textsf{R}}{}}

\makeatletter
\newcommand{\distas}[1]{\mathbin{\overset{#1}{\kern\z@\sim}}}%
\newsavebox{\mybox}\newsavebox{\mysim}
\newcommand{\distras}[1]{%
  \savebox{\mybox}{\hbox{\kern3pt$\scriptstyle#1$\kern3pt}}%
  \savebox{\mysim}{\hbox{$\sim$}}%
  \mathbin{\overset{#1}{\kern\z@\resizebox{\wd\mybox}{\ht\mysim}{$\sim$}}}%
}
\makeatother


\title[A new method for clustered survival data]{A new method for clustered survival data: Estimation of treatment effect heterogeneity and variable selection}

\author[1]{Liangyuan Hu\footnote{Corresponding author: {\sf{e-mail: liangyuan.hu@rutgers.edu}}, Phone: 732-235-4646, Fax: 732-235-5476}\inst{,1} }
\address[\inst{1}]{Department of Biostatistics and Epidemiology, Rutgers University, Piscataway, New Jersey 08854}

\Receiveddate{zzz} \Reviseddate{zzz} \Accepteddate{zzz} 

\begin{document}

 \begin{abstract}
We recently developed a new method riAFT-BART to draw causal inferences about population treatment effect on patient survival from clustered and censored survival data while accounting for the multilevel data structure. The practical utility of this method goes beyond the estimation of population average treatment effect. In this work, we exposit how riAFT-BART can be used to solve two important statistical questions with clustered survival data: estimating the treatment effect heterogeneity and variable selection. Leveraging the likelihood-based machine learning, we describe a way in which we can draw posterior samples of the individual survival treatment effect from riAFT-BART model runs, and use the drawn posterior samples to perform an exploratory treatment effect heterogeneity analysis to identify subpopulations who may experience differential treatment effects than population average effects.  There is sparse literature on methods for variable selection among clustered and censored survival data, particularly ones using flexible modeling techniques. We propose a permutation based approach using the predictor's variable inclusion proportion  supplied by the riAFT-BART model for variable selection. To address the missing data issue frequently encountered in health databases, we propose a strategy to combine bootstrap imputation and riAFT-BART for variable selection among incomplete clustered survival data. We conduct an expansive simulation study to examine the practical operating characteristics of our proposed methods, and provide empirical evidence that our proposed methods perform better than several existing methods across a wide range of data scenarios. Finally, we demonstrate the methods via a case study of predictors for in-hospital mortality among severe COVID-19 patients and estimating the heterogeneous treatment effects of  three COVID-specific medications. The methods developed in this work are readily available in the $\R$ package $\textsf{riAFTBART}$.
\end{abstract}

\maketitle

\section{Introduction}
Large-scale healthcare datasets garnered from multiple facilities are increasingly available, offering fertile ground for innovative investigation into emerging medical research questions. The inherent hierarchical data structure, in conjunction with the censored survival outcome, which is of strong clinical interest, poses considerable challenges for statistical analysis.  \cite{hu2022flexible} recently developed a flexible approach, riAFT-BART to estimate the population average treatment effect on patient survival, while accounting for the clustering structure of the observations. The method riAFT-BART stands for random-intercept accelerated failure time model with Bayesian additive regression trees (BART) \citep{chipman2010bart}. The utility of this new method, which leverages Bayesian machine learning, goes beyond the estimation of population average treatment effect. In this article, we exposit how riAFT-BART can be used to solve two important statistical questions with clustered survival data:  estimating the treatment effect heterogeneity and variable selection.  

An immediate extension of the new method, riAFT-BART, beyond the population average treatment effect is the estimation of  heterogeneous treatment effect. Prior work \citep{hu2021estimating} has shown that compared to algorithm-based machine learning, leveraging the likelihood-based machine learning technique BART can provide more accurate estimation of the individual survival treatment effect, which in turn can be used to assess the treatment effect heterogeneity. To the best of our knowledge, no work has yet discussed how BART can be used to estimate the heterogeneous effect of multiple treatments on patient survival using clustered survival observations.

Another important utility of riAFT-BART is for variable selection with clustered survival data, on which there is sparse literature. \cite{fan2002variable} proposed a family of variable selection methods for the gamma frailty model based on a nonconcave penalized likelihood approach using the Newton-Raphson method. \cite{androulakis2012estimation} extended the penalized Cox or Gamma frailty model to a  class of frailty models and proposed a generalized form of the likelihood function to allow the use of different continuous distributions for the frailty term. \cite{utazirubanda2021variable} proposed group LASSO with gamma-distributed frailty for variable selection in Cox regression. However, no corresponding software is available for these methods. 
\cite{bender2018generalized} proposed a piece-wise exponential additive mixed model (PEAMM) for clustered survival data, 
and \cite{marra2011practical} described a practical approach by which shrinkage penalties can be added to the smoothing parameters in PEAMM and variable selection can be performed by using  restricted maximum likelihood estimation.  \cite{rondeau2003maximum} proposed to conduct variable selection by maximizing the penalized likelihood using the Marquardt algorithm \citep{marquardt1963algorithm}, which is a combination of the Newton-Raphson algorithm and the steepest descent algorithm. \cite{ha2014variable} proposed a penalized h-likelihood estimation approach for variable selection of fixed effects using semiparametric frailty models.   These methods rely on parametric assumptions about the exact relationships between covariates and survival outcomes. Misspecifying  the parametric forms of  covariate-outcome relationships can reverberate through the variable selection procedure and yield unsatisfactory results \citep{hu2021variable,bleich2014variable}. 

Flexible machine learning modeling techniques can help relax the parametric assumptions and improve variable selection results.  \cite{ishwaran2010high} derived the distribution of the minimal depth of a maximal subtree, which measures the predictiveness of a variable in a survival tree, and used it for high-dimensional variable selection using random survival forests. 
\cite{lagani2010structure} proposed an algorithm based on the theory of Bayesian networks and the Markov blanket for variable selection suitable for high-dimensional and right-censored survival data. However, neither of these two machine learning based methods accommodates clustered survival observations.

In this work, we describe ways in which riAFT-BART can be used to (i) estimate the treatment effect heterogeneity, and to (ii) select important predictors in the presence of missing data with clustered survival observations.  The rest of the paper is organized as follows. Section~\ref{sec:TEH} describes the use of riAFT-BART for estimating individual survival treatment effects and exploring subgroups who may experience enhanced or reduced treatment effect than population average treatment effect.  Section~\ref{sec:VS} proposes a variable selection method using riAFT-BART when covariates are subject to missing data.  
In Section~\ref{sec:sim}, we conduct an expansive simulation to assess the practical operating characteristics of riAFT-BART in estimating treatment effect heterogeneity and performing variable selection. Section~\ref{sec:application} illustrates the methods using a COVID-19 dataset drawn from the Mount Sinai Medical Center, and Section~\ref{sec:discussion} provides a discussion.

\section{Estimation of Treatment Effect Heterogeneity} \label{sec:TEH}
\subsection{Notation}\label{sec:notation}
We maintain notation used in  \cite{hu2022flexible}. 
Consider a nonexperimental study with  $K$ clusters, each having treated $n_k$ individuals, indexed by $i=1, \ldots, n_k, k=1, \ldots, K$. The total sample size is $N = \sum_{k=1}^K n_k$. There are $J$ possible treatment options, denoted by $A \in \mathscr{A} = \{a_1, \ldots,a_J\}$.  Denote pre-treatment measured, $L$-dimensional covariates by $\bm{X}_{ik}=\{X_{ik1},\ldots, X_{ikL}\}$ for each individual $i$ in cluster $k$, whose failure time $T_{ik}$ may be right censored at $C_{ik}$. The observed outcome consists of $Y_{ik}=\min (T_{ik}, C_{ik})$ and the censoring indicator $\Delta_{ik}=I (T_{ik}<C_{ik})$. Let $V_k$ be the cluster indicators.  Proceeding with the counterfactual framework, the counterfactual failure time under treatment $a_j$ for individual $i$ in cluster $k$ is defined as $T_{ik} (a_j)$, $ \forall a_j \in \mathscr{A}$, and the counterfactual censoring time under treatment $a_j$ is defined as $C_{ik}(a_j)$. Throughout, we maintained the standard assumptions \citep{hu2022flexible,hu2021estimating,hu2019causal,chen2001causal,arpino2016propensity}: consistency, weak unconfoundedness, positivity and covariate-dependent censoring, for drawing causal inference with observational clustered survival data. Detailed assumptions and implications of assumptions can be found in \cite{hu2022flexible}.

\subsection{Method}
\cite{hu2022flexible} adapted BART into a random-intercept accelerated failure time model, 
\begin{eqnarray}
\begin{aligned}
\label{eq: riAFT-BART}
&\log T_{ik} = f(A_{ik}, \bm{X}_{ik}) + b_k + \epsilon_{ik}, \\
 b_k \distas {i.i.d} & N(0,\alpha_k \tau^2), \quad \epsilon_{ik} \distas {i.i.d} N(0,\sigma^2), \quad b_k \perp \epsilon_{ik}, 
\end{aligned}
\end{eqnarray} 
where $f(A_{ik}, \bm{X}_{ik})$ is an unspecified function relating treatment assignment $A_{ik}$ and pretreatment confounders $\bm{X}_{ik}$ to survival times $T_{ik}$,  $b_k$'s are the random intercepts capturing cluster-specific main effects, and $\epsilon_{ik}$ is the residual term. The unknown function $f$ is flexibly modeled by BART via a sum of shallow trees \citep{hu2020estimation,hu2021estimation}. A mean-zero normal distribution and independence are assumed for $b_{k}$ and $\epsilon_{ik}$. A redundant parameter $\alpha_k$ was used for the variance of $b_k$ as a parameter expansion  technique \citep{gelman2008using} to improve computational performance. Regularizing priors are placed on the tree parameters to keep the impact of each individual tree on the overall fit small and thus prevent overfitting \citep{chipman2010bart}. Sensible priors are used for parameters $\sigma^2$, $\tau^2$ and $\alpha_k$ \citep{hu2022flexible,henderson2020individualized}. To deal with right censoring, data augmentation is used. That is, the unobserved survival times are imputed from a truncated normal distribution in each Gibbs iteration. A Metropolis within Gibbs procedure \citep{hu2022flexible} (Web Section 1) was proposed to draw posterior inferences about the population average treatment effect on patient survival, which is defined as
$$\theta_{a_j, a_{j'}} = \frac{1}{N}\sum_{k=1}^K \sum_{i=1}^{n_k} E \lsq \log T_{ik}(a_j)- \log T_{ik}(a_{j'}) \cond \bm{X}_{ik} = \bm{x}_{ik}, V_k=v_k \rsq, \; \forall a_j, a_{j'} \in \mathscr{A}.$$
Via outcome modeling, the pairwise population average treatment effect between $a_j$ and $a_{j'}$ can be estimated as 
  \begin{align} \label{eq:theta_hat}
      \hat{\theta}_{a_j,a_{j'}} =  \frac{1}{D} \sum_{d=1}^D \frac{1}{N} \sum_{k=1}^K \sum_{i=1}^{n_k} \lbc \tilde{f}^d \lp a_j, \bm{x}_{ik}\rp - \tilde{f}^d \lp a_{j'},\bm{x}_{ik} \rp \rbc,
  \end{align}
  where $\tilde{f}^d$ is the $d$th draw from the posterior distribution of $f$. Inferences can be obtained based on the $D$ posterior average treatment effects, $(1/N) \sum_{k=1}^K\sum_{i=1}^{n_k}\lbc \tilde{f}^d \lp a_j, \bm{x}_{ik} \rp - \tilde{f}^d \lp a_{j'}, \bm{x}_{ik} \rp \rbc, d =1,\ldots, D$. 

Turning to the estimation of treatment effect heterogeneity. We first define the \emph{individual survival treatment effect} (ISTE)  for individual $i$ in cluster $k$ as 
\begin{eqnarray}\label{eq:ISTE}
\zeta_{a_j,a_{j'}}(\bm{x}_{ik})= E\lsq \log T_{ik}(a_j) - \log T_{ik}(a_{j'}) \cond \bm{x}_{ik}, v_k\rsq,
\end{eqnarray}
which will be used as the building block for estimating the heterogeneous treatment effect.  From the fitted Model~\eqref{eq: riAFT-BART}, we can estimate $\zeta_{a_j,a_{j'}}(\bm{x}_{ik})$ as
\begin{eqnarray*} 
 \hat{\zeta}_{a_j,a_{j'}}(\bm{x}_{ik}) = \tilde{f} (a_j, \bm{x}_{ik}) -\tilde{f} (a_{j'}, \bm{x}_{ik}), 
\end{eqnarray*}
where $\tilde{f}$ is the mean of $D$ draws from the posterior predictive distribution of $f$. The ISTE estimates $\{\hat{\zeta}_{a_j,a_{j'}}(\bm{x}_{ik}), i=1,\ldots, n_k, k=1, \ldots, K\}$ can then be used to explore the degree of treatment effect heterogeneity. By the ``virtual twins'' approach \citep{foster2011subgroup}, the  $\{\hat{\zeta}_{a_j,a_{j'}}(\bm{x}_{ik})\}$ would be regressed on the predictors via a tree diagram to identify the predictor subspaces that explain away differences in $\{\hat{\zeta}(\bm{x}_{ik})\}$. In a similar spirit, the ``fit-the-fit'' strategy has been used to identify possible subgroups defined by combination rules of covariates that have differential treatment effects \citep{logan2019decision,aoehu2021}. Mathematically, we wish to find a covariate subspace $\mathcal{S}_{a_j, a_{j'}} = \{S^1_{a_j, a_{j'}}, S^2_{a_j, a_{j'}}, \ldots, S^H_{a_j, a_{j'}}\}$, where $h \in \{1,\ldots,H\}$ is the number of combination rules of covariates, such that the average treatment effects across individuals in this region are considerably different than the population average treatment effect. Denote the measure of subgroup treatment effect between $a_j$ and $a_{j'}$ as $\mathcal{M}(\mathcal{S}_{a_j, a_{j'}}) = \{M (S^1_{a_j, a_{j'}}), \ldots, M (S^H_{a_j, a_{j'}})\}$, and
\begin{eqnarray} \label{eq:MS-subg}
M(S^h_{a_j, a_{j'}}) = E \lsq \zeta_{a_j,a_{j'}}(\bm{x}_{ik}) \cond  \bm{x}_{ik} \in S^h_{a_j, a_{j'}} \rsq .
\end{eqnarray}
We adopt the ``fit-the-fit'' strategy to estimate $\mathcal{S}_{a_j,a_{j'}}$ and $\mathcal{M}(\mathcal{S}_{a_j,a_{j'}})$. The algorithm proceeds with the following steps: 
\begin{enumerate}
\item  Add each candidate predictor separately to the random forests model \citep{breiman2001random} using $\{\hat{\zeta}_{a_j,a_{j'}}(\bm{x}_{ik})\}$ as the outcome, and record the model fit measured by  $R^2$; include the predictor  corresponding to the largest $R^2$ in the model;
\item Add each of the remaining candidate predictors separately to the previous step's model and retain the predictor corresponding to the largest $R^2$ improvement;
\item  Repeat step 2 until the percentage improvement in $R^2$ is less than 1\%. At this point, the final model is established.
\end{enumerate}
The $R^2$ is the coefficient of determination in regression analysis that determines the proportion of variance in the dependent variable that can be predicted from the independent variable(s). Different from prior work, we use random forests instead of a single tree to regress the ISTE estimates on predictors for improved modeling accuracy and stability. To interpret the tree ensembles in the final random forests model yielded in Step 3, we apply the $\code{inTrees}$ algorithms \citep{deng2019interpreting} to translate the composite rules into a simple rule model. The $\code{inTrees}$ algorithms extract, measure and rank rules, prune redundant rules, detect frequent variable interactions, and summarize rules into a simple tree diagram model that can be used for predicting new data. The branch decision rules in the final tree diagram model sending individuals to the terminal nodes suggest the predictor subspace $\hat{\mathcal{S}}_{a_j, a_{j'}}$; and subgroup treatment effects that are substantially different from the population effect $\theta_{a_j,a_{j'}}$ can then be estimated by averaging the ISTE estimates among individuals falling into each of terminal nodes, 
$$\hat{M}(\hat{S}^h_{a_j, a_{j'}}) =\frac{1}{\cond \hat{S}^h_{a_j,a_{j'}}\cond} \sum \limits_{i,k:\bm{x}_{ik} \in \hat{S}^h_{a_j,a_{j'}}} \hat{\zeta}_{a_j,a_{j'}}(\bm{x}_{ik}),$$ 
where $\mid \hat{S}^h_{a_j,a_{j'}} \mid$ is the size of  $\hat{S}^h_{a_j,a_{j'}}$.

\section{Variable Selection}\label{sec:VS}

Variable selection has been a longstanding statistical problem and  studied from both the classical and Bayesian perspective. In the field of causal inference, variable selection methods can aid in 
confounder selection \citep{vanderweele2019principles}. Variable selection methods are largely based on parametric models, in which the exact relationships between the response and predictor variables need to be explicitly specified.  However, incorrectly specifying the parametric models may produce unsatisfactory variable selection results \citep{bleich2014variable,hu2021variable,lin2022a}.  For example, noise variables may be selected and important variables ignored. 
Flexible machine learning models can better represent functional forms of arbitrary complexity, thus mitigate  parametric assumptions and attendant errors in variable selection. We develop a way in which the new method riAFT-BART can be used for variable selection among clustered survival data. Notation defined in Section~\ref{sec:notation} is largely retained with the exception of treatment assignment $A_{ik}$, as the focus now is variable selection. 

\subsection{Random-intercept accelerated failure time model with Bayesian additive regression trees}
We propose a permutation-based approach for variable selection. Following prior work, \cite{bleich2014variable}, \cite{hu2021variable}, and \cite{lin2022a}, we use the variable inclusion proportion (VIP) of each predictor variable as the foundation to determine how important a predictor variable is.  The VIP score is supplied by the BART model.  The VIP score signifies the frequency of each predictor's utilization in defining the splitting rules, relative to the total count of such rules appearing in the model. Each predictor's proportion can be used to rank the variables based on their relative importance. However, these proportions should not be interpreted as prescriptive guidelines for variable selection. Instead, they serve as indicative measures of variable significance within the model \citep{chipman2010bart,hu2020ranking}.
We note that in the riAFT-BART model~\eqref{eq: riAFT-BART}, the predictor variables $\bm{X}_{ik}$, along with treatment assignment $A_{ik}$, is included in the unknown function $f(A_{ik}, \bm{X}_{ik})$, which is being modeled by BART. Thus, in each iteration of the riAFT-BART sampling algorithm \citep{hu2022flexible}, we can extract the VIP for each predictor variable in the step where a BART model is fit to regress $\log y^{cent,c}_{ik} - b_k$ on $\{A_{ik}, \bm{X}_{ik}\}$, where $y^{cent,c}_{ik}$ denotes the centered and complete  survival time. Note that right censoring is addressed via data augmentation. Detailed riAFT-BART sampling algorithm appears in Web Section 1. For each predictor variable $X_l, \; l\in \{1, \ldots, L\}$ (we suppress subscripts $i$ and $k$ for brevity of notation), we compute its VIP by averaging the VIP scores across $D$ posterior draws \citep{bleich2014variable},  $\text{VIP}_{x_l} = \frac{1}{D} \sum_{d=1}^D \widetilde{\text{VIP}}^d_{x_l}, d =1,\ldots, D$, where $\widetilde{\text{VIP}}^d_{x_l}$ is the $d$th draw from the posterior distribution of the VIP for $X_l$. In our simulation (Section~\ref{sec:sim}), we set $D = 4500$ with the first 1000 discarded as burn-in. Trace plot was used to check the convergence of the posterior distribution of the VIP for $X_l$.

Building on prior literature \citep{bleich2014variable, huang2010variable}, we propose a non-parametric and permutation-based approach for variable selection, by which we first permute the event times together with the censoring indicators and then establish  the thresholds for variable selection using the VIP from observed data and the permuted data. Specifically, we first create $P$ permutations of the vectors of observed survival times and event indicators $(\bm{y}, \bm{\Delta}) = \{Y_{11}, \Delta_{11}, \ldots, Y_{N}, \Delta_{N}\}$: $(\bm{y}_1^*, \bm{\Delta}_1^*), (\bm{y}_2^*, \bm{\Delta}_2^*) \ldots, (\bm{y}_P^*, \bm{\Delta}_P^*)$. Then for each of the permuted outcome vectors $(\bm{y}_p^*, \bm{\Delta}_p^*)$, we fit an riAFT-BART model with $(\bm{y}_p^*, \bm{\Delta}_p^*)$ as the response and the original  $\{\bm{X}_{ik}\}$ as predictor variables and $V_k$ as cluster indicator. 
 This permutation strategy removes any dependency between the predictors $\{\bm{X}_{ik}\}$ and the outcomes  $\{(Y_{ik}, \Delta_{ik})\}$ while keeping possible dependencies among the predictor variables. 
From the riAFT-BART run using each permuted outcome vector $(\bm{y}_p^*, \bm{\Delta}_p^*)$,  we retain the VIP for each predictor variable. 
We use $\text{VIP}_{x_l,p}^*$ to denote the VIP score from riAFT-BART for predictor $X_l$ from the $p$th permuted outcomes, and 
use the notation $\text{VIP}_{\bm{X},p}^*$ for the vector of all $L$ VIPs from the $p$th permuted outcomes,  $(\text{VIP}_{x_1,p}^*, \ldots, \text{VIP}_{x_L,p}^*)$. As suggested by \cite{bleich2014variable}, we use the VIPs across all $P$ permutations $\text{VIP}^*_{\bm{X},1}, \ldots, \text{VIP}^*_{\bm{X},P}$ as the ``null'' distribution for the VIPs from the unpermuted outcomes $(\bm{y},\bm{\Delta})$, $\text{VIP}_{x_1}, \ldots, \text{VIP}_{x_L}$. The permutation null distribution will be used to determine which variables should be selected or deemed as important. 
We adopt the ``local'' threshold strategy, which has been shown in several works \citep{bleich2014variable, hu2020tree,hu2021variable,lin2022a} to have better performance than more stringent strategies such as ``global max'' for large-scale healthcare data in which the ratio of sample size and number of predictors is relatively large. More stringent strategies compare  the VIP of a predictor $\text{VIP}_{x_l}$ to the permutation distribution across all predictor variables $\text{VIP}_{\bm{X},1}^*, \ldots, \text{VIP}_{\bm{X},P}^*$.  By contrast, a local threshold compares $\text{VIP}_{x_l}$ with its own permutation null distribution $\text{VIP}_{x_l,1}^*, \ldots, \text{VIP}_{x_l,P}^*$. Specifically, we only select predictor $X_l$ if $\text{VIP}_{x_l}$ is greater than the $1-\alpha$ quantile of the
permutation distribution $\text{VIP}_{x_l,1}^*, \ldots, \text{VIP}_{x_l,P}^*$. Following prior work \citep{hu2021variable,lin2022a}, we set $\alpha=0.05$ in our simulation and case study, because this threshold value has been shown to yield good variable selection results with a balance between selecting useful and selecting noise predictors. 

\subsection{Comparison methods}
\subsubsection{Piece-wise Exponential Additive Mixed Model} 
For a PEAMM \citep{bender2018generalized}, the hazard rate at time $t$  for individual $i$ in cluster $k$  with covariates $\bm{X_{ik}}$ and cluster indicator $V_k$ is given by
\begin{eqnarray} \label{eq:peamm}
\lambda \lp t \cond \bm{X}_{ik}, V_k \rp = \exp \lbc g_0 \lp t_q \rp + \sum_{m=1}^{M}g_m \lp \bm{x}_{ik,m} \rp + b_k \rbc \forall t \in (\kappa_{q-1}, \kappa_{q}),
\end{eqnarray}
where $g_0 \lp t_q \rp$ represents the log-baseline hazard rate, $g_m \lp \bm{x}_{ik,m} \rp$ is the basis functions of natural cubic splines associated with the covariates $\bm{x}_{ik}$,  $b_k$ is the random intercept term capturing the cluster-specific effect, and $\kappa_{q-1}$ and $\kappa_{q}$ are the start and stop times of the censored survival dataset represented in the (start, stop] counting process form. 
For variable selection using PEAMM, following the practical approach by \cite{marra2011practical},  a shrinkage penalty term can be added to each basis term $g_m \lp \bm{x}_{ik,m} \rp$ of model~\eqref{eq:peamm}: 
\begin{eqnarray*}
 \lambda \lp t \cond \bm{X}_{ik}, V_k \rp = \exp \lbc g_0 \lp t_q \rp + \sum_{m=1}^{M}
\lp g_m \lp \bm{x}_{ik,m} \rp + \int g'_m(\bm{x}_{ik,m})^2d\bm{x}_{ik,m} \rp + b_k \rbc \forall t \in (\kappa_{q-1}, \kappa_{q}),
\end{eqnarray*}
so that when  the whole spline basis terms for a predictor $x_{ikl}$ are shrunk to zero, the variable $x_{ikl}$ will be selected out of the model. The shrinkage penalties, along with all other model coefficients, can be estimated using restricted maximum likelihood estimation. 

\subsubsection{Frailty Models by Regularization Methods}

Given the common unobserved frailty $u_{k}$ for cluster $k$, the conditional hazard function at time $t$ for individual $i$ in cluster $k$ is
\begin{eqnarray}\label{eq:frailty_mod}
\lambda(t \cond \bm{X}_{ik}, V_k) = u_{k} \lambda_0(t)\exp(\bm{x_{ik}}^T\bm{\beta})
\end{eqnarray}
where $\lambda_0(t)$ is an unspecified baseline hazard function and $\bm{\beta} = (\beta_1, \beta_2, ..., \beta_L)^T$ is a vector of regression parameters for  $\bm{x_{ik}}$. The common unobserved frailty $u_{k}$ is assumed to follow a gamma distribution because of its mathematical convenience. \cite{rondeau2003maximum} proposed to conduct variable selection by maximizing the penalized likelihood,
$$
l(\lambda(t \cond \bm{X}_{ik}, V_k))- \alpha \int_{0}^{\infty} \lambda_{0}''^2 (t)dt,
$$
where $l(\lambda(t \cond \bm{X}_{ik}, V_k))$ is the full marginal log-likelihood for the frailty model~\eqref{eq:frailty_mod} and $\alpha$ is the  smoothing parameter which controls the trade-off between the data fit and the smoothness of the baseline hazard function.  
The regression parameters can be estimated by the robust Marquardt algorithm \citep{marquardt1963algorithm} -- a combination of the Newton Raphson algorithm and the steepest descent algorithm.  
\cite{ha2014variable} instead proposed to perform variable selection by maximizing the penalized profile h-likelihood,
$$
l^*(\lambda(t \cond \bm{X}_{ik}, V_k)) - \sum_{l=1}^{L} B_\gamma(|\beta_l|),
$$
where $l^*(\lambda(t \cond \bm{X}_{ik}, V_k))$ is the profiled h-likelihood of the frailty model, $B_\gamma(|\beta_l|)$ is a penalty function that controls model complexity using the tuning parameter $\gamma$, with the optimal value determined by the Bayesian information criterion (BIC).  Three penalty functions were considered for $B_\gamma(|\beta_l|)$: least absolute shrinkage and selection operator \citep{tibshirani1996regression}, smoothly clipped absolute deviation \citep{fan2001variable}, and h-likelihood \citep{lee2014new}.  

\subsubsection{Backward stepwise selection} 
We also implement a classical variable selection approach through a recursive procedure of backward stepwise selection based on statistical testing, as described in \cite{wood2008should} The backward stepwise selection starts with a random-intercept Cox regression model \citep{cortinas2005version} with all potential predictors and removes from the model the predictor that has the least impact on the fit determined by the $\alpha$ significance level. The forward selection process then checks whether removed variables should be added back into the model at $\alpha (1-\epsilon)$ for $\epsilon$ small significance level. We use $\alpha=0.05$ as suggested by \cite{wood2008should}.

\subsection{Variable selection with missing data} \label{sec:vs-mis}
Large-scale healthcare data are susceptible to the missing data issue. Missing values in  predictor variables pose challenges for variable selection.  \cite{long2015variable} proposed a general
resampling approach that combines bootstrap imputation and randomized lasso based variable selection procedure, and
used simulations to demonstrate that this approach had better performance compared with several existing methods for variable selection in the presence of missing data.  Recent work by \cite{hu2021variable} shows that combining bootstrap imputation with flexible machine learning based variable selection methods can substantially improve variable selection results over methods that rely on parametric modeling of the covariate-outcome relationship. In this work, we combine bootstrap imputation and variable selection, and compare each of the methods considered on the ability to select
 predictor variables that are truly associated with the outcome, or useful predictor variables, among \emph{incomplete} clustered survival data. 
  To ensure consistency with previous literature, such as \cite{wood2008should}, 
  \cite{bleich2014variable}, and \cite{hu2021variable}, 
  and to facilitate  comparison of our methods and results with similar studies, we adopt the four performance metrics frequently employed in the variable selection literature. A good method should have high values of precision, recall and $F_1$ , and low Type I error.

 \begin{enumerate}[label={(\arabic*)}]
     \item  $\text{Precision} =\text{TP}/ (\text{TP} + \text{FP})$, where TP and FP are respectively  the number of true positive and false positive selections.  A true positive selection is when a useful predictor was selected; and a false positive selection is when a noise predictor was selected.  The \emph{precision} of a variable selection method is the proportion of truly useful predictors among all selected predictors. 
     \item  $\text{Recall} =\text{TP}/(\text{TP}+\text{FN})$, where FN is the number of false negative selections. A false negative selection is when a truly useful predictor was not selected. The \emph{recall} of a variable selection method is the proportion of truly useful variables selected among all useful variables. This is sometimes referred to as the \emph{power} in the literature \citep{wood2008should}.
     \item $F_1 =  2\text{ Precision} \times \text{Recall} / (\text{Precision} + \text{Recall}).$ The $F_1$ score measures a method's ability to avoid selecting irrelevant predictors (precision) with its ability to identify the full set of useful predictors (recall). 
     \item Type I error. The \emph{Type I error} measures the overall probability of a method incorrectly selecting the noise predictor -- the mean of the probabilities that the method  incorrectly selects each of the noise predictors. 
 \end{enumerate}


Among incomplete data, we first conduct bootstrap imputation through two steps: (i) generate $B$ bootstrap datasets of the original data; (ii) conduct a single imputation for each bootstrap dataset using an imputation method of choice. Following \cite{hu2021variable}, in our simulations and  case study, we used $B=100$ and performed  standard imputation program using the $\R$ package $\code{mice}$. Variable selection via each of the methods considered was performed for all 
$B$ bootstrap imputed datasets, and the final set of predictors were selected if they were selected in at least $\pi B$ imputed datasets, where $\pi \in (0,1)$ is a fraction threshold for selecting a predictor.

\section{Simulation} \label{sec:sim}

\subsection{Treatment effect heterogeneity}
We carried out a contextualized simulation to evaluate the performance of our proposed method in estimating the heterogeneous treatment effect using observational data with clustered survival outcomes and multiple treatments. For methods comparison, we adapted the popularly used inverse probability weighting method into the setting of clustered and censored survival data to form two comparison methods: inverse probability of treatment weighting with the random-intercept Cox regression model (IPW-riCox) and doubly robust random-intercept additive hazards model (DR-riAH) \citep{cai2011additive}. Additionally, we considered the PEAMM \citep{bender2018generalized}  --  another outcome modeling based method  flexible at capturing nonlinear relationships. Because not all methods directly model the survival times $T_{ik}$, to objectively compare methods, we use the individual-specific counterfactual survival curve as the basis. That is, the ISTE $\zeta_{a_j, a_{j'}} (\bm{x}_{ik})$  will now be defined in a similar way as in equation~\eqref{eq:ISTE} but using functionals of the counterfactual survival curve -- either 
 the survival probability at a fixed point in time  or the conditional restricted mean survival time (RMST) \citep{royston2013restricted}. We will compare each method's ability to accurately estimate the ISTE  $\zeta_{a_j, a_{j'}} (\bm{x}_{ik})$ and the average subgroup treatment effect  $M(S^h_{a_j,a_{j'}})$.  The performance metrics used to evaluate the methods were the relative bias (or percent bias), root-mean-squared-error (RMSE) and precision in the estimation of heterogeneous effects (PEHE) \citep{hu2021estimating}. The relative bias evaluates the bias in the estimation of treatment effect on the relative scale with respect to the true treatment effect (which is constant across different methods). The PEHE measures the difference between the true and estimated survival probabilities or RMST across all data points. A smaller value of PEHE indicates a higher estimation accuracy. Although the methods examined in this study rely on different models, each can be utilized to estimate survival probabilities and RMST. These common outcomes will serve as a basis for an impartial comparison of these methods.

For weighting based  methods IPW-riCox and DR-riAH, we used Super Leaner \citep{van2007super} to estimate the stabilized inverse probability of treatment weights. Super Learner was implemented using the $\R$ package \code{SuperLearner} and the library argument was set to \code{SL.library = c("SL.xgboost", "SL.bartMachine","SL.gbm")}. We used  $\R$ package \code{coxme} to implement IPW-riCox and \code{timereg} to implement DR-riAH.  For PEAMM, the \code{gam} function from $\R$ package \code{mgcv} and  two helper functions, (\code{as\_ped} and \code{add\_surv\_prob}), from $\R$ package \code{pammtools} were used. 
For all methods, all confounding variables available to the analyst in their original forms were included in the corresponding models. 


\subsubsection{Simulation design} \label{sec:sim-design-teh}
We simulated datasets with structures similar to the COVID-19 data used in our case study (Section~\ref{sec:application}).  We generated $K=10$ clusters, each with a sample size of $n_k=200$. The total sample size is $N=2000$. We simulated 7 confounding variables for each individual $i$ in cluster $k$: five continuous variables  $X_{ikl} \sim N(0,1), l \in  \{1,\ldots,5\}$; and two categorical variables  $X_{ikl} \sim Multinomial(1,.3,.3,.4), l=6,7$. We generated $J=3$ treatment groups with unequal sample sizes; the ratio of individuals across treatment groups was 5:3:1, which is similar to the ratio of individuals across the three treatment groups in our case study. The treatment assignment mechanism follows a random-intercept multinomial logistic regression model,
\begin{equation*}\label{eq:trt_assign}
\begin{split}
\ln  \dfrac{P(A_{ik}=1)}{P(A_{ik}=3)} &= 1.5+.1X_{ik1}+.1X_{ik2} +.1X_{ik3} + .5X_{ik4} + .4X_{ik5} +.2X_{ik6} + .3X_{ik7} \\
& +.4X_{ik2}^2+ .4X_{ik2}^2X_{ik5} + \tau_k \\
\ln  \dfrac{P(A_{ik}=2)}{P(A_{ik}=3)} &= .7 +.1X_{ik1}+.3X_{ik2} +.2X_{ik3} + .2X_{ik4} + .1X_{ik5} +.4X_{ik6} + .5X_{ik7} \\
&-.3X_{ik2}X_{ik4}+ .7X_{ik2}^2X_{ik4} + \tau_k,
\end{split}
\end{equation*}
where the random intercept $\tau_k \sim N(0,1^2)$. The parameter setup induces  a moderate level of confounding, leading to moderate covariate overlap. The sparsity of overlap can be visualized by the distributions of true generalized propensity scores, shown in Web Figure 1.  It should be noted that the overall extent of covariate overlap closely resembles that observed in the COVID-19 dataset (refer to Web Figure 8). The specific one-to-one correspondences are as follows: $A=1$ emulates treatment 1, dexamethasone ; $A=3$ emulates treatment 2, remdesivir; and $A=2$ emulates treatment 3, a combination of dexamethasone and remdesivir. Following strategies \citep{hu2021estimating, lu2018estimating} recommended for assessing the performance of methods in estimating heterogeneous treatment effects, we subclassified the simulated individuals into 40 subgroups based on the distribution of true generalized propensity scores \citep{feng2012generalized}, and calculated the biases and RMSE for each subgroup across 250 data replications. Note that the generalized propensity scores for each individual sum to one,  $P(A_{ik}=1\mid \bm{X}_{ik},\tau_k )+ P(A_{ik}=2 \mid \bm{X}_{ik},\tau_k )+P(A_{ik}=3 \mid \bm{X}_{ik},\tau_k )=1$.  The subclassification is based on intervals of the true generalized propensity scores for treatment 1 and treatment 2, representing a full range of assignment propensity to each treatment group.  Subclasses of generalized propensity scores are presented in Web Table 1.

We simulated three sets of true counterfactual survival times from the Weibull survival distribution, which can be parameterised as both the accelerated failure time model and  Cox proportional hazards model \citep{hu2022flexible},
\begin{eqnarray*}
 S_{ik}(t) = \exp \lsq -\lbc\lambda_{a_j}\exp\lp q_{a_j}\lp \bm{X}_{ik},b_k \rp \rp t^{\eta}\rbc  \rsq,
\end{eqnarray*}
where $S_{ik}(t)$ is individual-specific survival function for $i$th individual in cluster $k$, $\lambda_{a_j}$ is a treatment group specific parameter, $\eta$ is the shape parameter and $q_{a_j}\lp \bm{X}_{ik}, b_k \rp$ represents a general functional form of covariates and the random-intercept. Using the inverse transform sampling, we generated counterfactual survival times by

\begin{eqnarray}
T_{ik}\lp a_j \rp = \lbc \frac{-\log U}{\lambda_{a_j}\exp\lp q_{a_j}\lp \bm{X}_{ik},b_k \rp \rp} \rbc ^{1/\eta}
\end{eqnarray}
for $a_j \in \{1,2,3\}$, where $U$ is a random variable following the uniform distribution on the interval $[0,1]$ and $\lambda_{a_j} = \{5000,800,1200\}$ for  
 $a_j = 1,2,3$.  Observed and uncensored survival times are generated as $$T_{ik} = \sum\limits_{a_j \in \{1,2,3\}} T_{ik}(a_j)I(A_{ik}= a_j).$$

 We considered three treatment effect heterogeneity settings with an increasing level of complexity:
\begin{enumerate}\small
 \item [(a)]  $q_1\lp \bm{X}_{ik},b_k \rp = .1X_{ik1} + .3X_{ik2}+ \sin(3.14X_{ik3})+ .6X_{ik4}+.5X_{ik5}+1.2X_{ik6} + .4X_{ik7} +.3X_{ik2}^2+.5X_{ik4}X_{ik5} +b_k-1$ 
 
 $q_2\lp \bm{X}_{ik},b_k \rp = .4X_{ik1} + 1.2\sin(3.14 X_{ik2})+.4X_{ik3}+.3X_{ik4}+1.0X_{ik5}+.8X_{ik6}+ .2X_{ik7}+.7X_{ik1}^2+.4X_{ik1}X_{ik4}+b_k $
 
 $q_3\lp \bm{X}_{ik},b_k \rp = .4X_{ik1} + .9X_{ik2} + .4X_{ik3} + .9X_{ik4}+.4X_{ik5}+.4X_{ik6}+ .3X_{ik7} +b_k -2$
 
 \item [(b)] $q_1\lp \bm{X}_{ik},b_k \rp = .1  X_{ik1} + .3X_{ik2}+ \sin(3.14 X_{ik3})+ .6X_{ik4}+.5X_{ik5}+1.2X_{ik6}+ .3X_{ik7} +.3X_{ik2}^2+.5X_{ik4}X_{ik5} +b_k-1$ 
 
 $q_2\lp \bm{X}_{ik},b_k\rp = .4X_{ik1} + 1.2\sin(3.14  X_{ik2})+.4X_{ik3}+.3X_{ik4}+1.0X_{ik5}+.8X_{ik6}+ .1X_{ik7}+.7X_{ik1}^2+.4X_{ik1}X_{ik4}+b_k$
           
$q_3\lp \bm{X}_{ik},b_k \rp = .4\sin(3.14 X_{ik1}) + .9X_{ik2}+ .9X_{ik3}+.4X_{ik4}+.4X_{ik5}+.9X_{ik6}+ .3X_{ik7}+.4X_{ik4}^2-.3X_{ik2}X_{ik3} +b_k -3$ 

\item [(c)]$q_1\lp \bm{X}_{ik},b_k \rp = .1  X_{ik1} + .3X_{ik2}+ \sin(3.14 X_{ik3})+ .6X_{ik4}+.5X_{ik5}+1.2X_{ik6} +.3X_{ik2}^2+.5X_{ik4}X_{ik5} +b_k-1$

$q_2\lp \bm{X}_{ik},b_k \rp = .4X_{ik1} + 1.2\sin(3.14 X_{ik3})+.4X_{ik4}+.3X_{ik5}+1.0X_{ik6}+.8X_{ik7}+.7X_{ik1}^2+.4X_{ik1}X_{ik4}+b_k $
               
$q_3\lp \bm{X}_{ik},b_k \rp = .4\sin(3.14 X_{ik2}) + .9X_{ik3}+ .9X_{ik4}+.4X_{ik5}+.4X_{ik6}+.9X_{ik7}+.4X_{ik4}^2-.3X_{ik2}X_{ik3} +b_k-3$,
 \end{enumerate}
 where the random intercept $b_k \sim N(0,4^2)$.  Across all three heterogeneity settings, $X_{ik3}, X_{ik4}, X_{ik5}, X_{ik6}$  were confounders related to both the treatment assignment mechanism and the potential outcome generating process. In scenario (a), nonlinear covariate effects are present in  treatment groups 1 and 2. In scenario (b), there are nonlinear covariate effects  in all three treatment groups. Scenario (c) is similar to scenario (b) except that a nonoverlapping set of covariates are included in three treatment groups. This additional complexity represents an additional source of treatment effect heterogeneity. The parameter $\eta$ was set to 2 and $\exp(0.7+0.5x_{ik1})$ to respectively induce proportional hazards  and nonproportional hazards.  We further generate the censoring time $C$ independently from an exponential distribution with the rate parameter chosen to induce a censoring proportion of 50\%, which is close to the censoring proportion observed in the COVID-19 data in our case study. The Kaplan-Meier survival curves by treatment group for all three heterogeneity settings under both proportional and nonprotional hazards are presented in Web Figure 2.

\subsubsection{Simulation results}

Table~\ref{tab:PEHE-survival-moderate} summarizes the overall precision measure PEHE (based on 3-week survival probability) for each of four methods under three heterogeneity settings and both proportional and nonproportional hazards. First, flexible methods riAFT-BART and PEAMM and doubly robust weighting method DR-riAH yielded substantially smaller PEHE than the parametric single robust method IPW-riCox across all simulation scenarios. Second,
as the heterogeneity setting became more
complex, flexible methods riAFT-BART and PEAMM tended to have better performance evidenced by decreasing mean and variation of the PEHE.
 All methods show decreasing precision in estimating ISTE, demonstrated by increasing PEHE value, when the proportional hazards assumption fails to hold.  This is possibly due to more complex data generating processes, in which the rate parameter $\eta$ depends on covariate $x_{ik1}$. The decrease in estimation precision is the smallest for riAFT-BART. Overall, across all scenarios, the proposed method riAFT-BART had the smallest PEHE, or the highest accuracy in estimating the ISTE.  Similar observations can be made from the PEHE measures based on 3-week RMST, which are provided in Web Table 2.

\begin{figure}[!ht]
    \centering
    \includegraphics[width=1\textwidth]{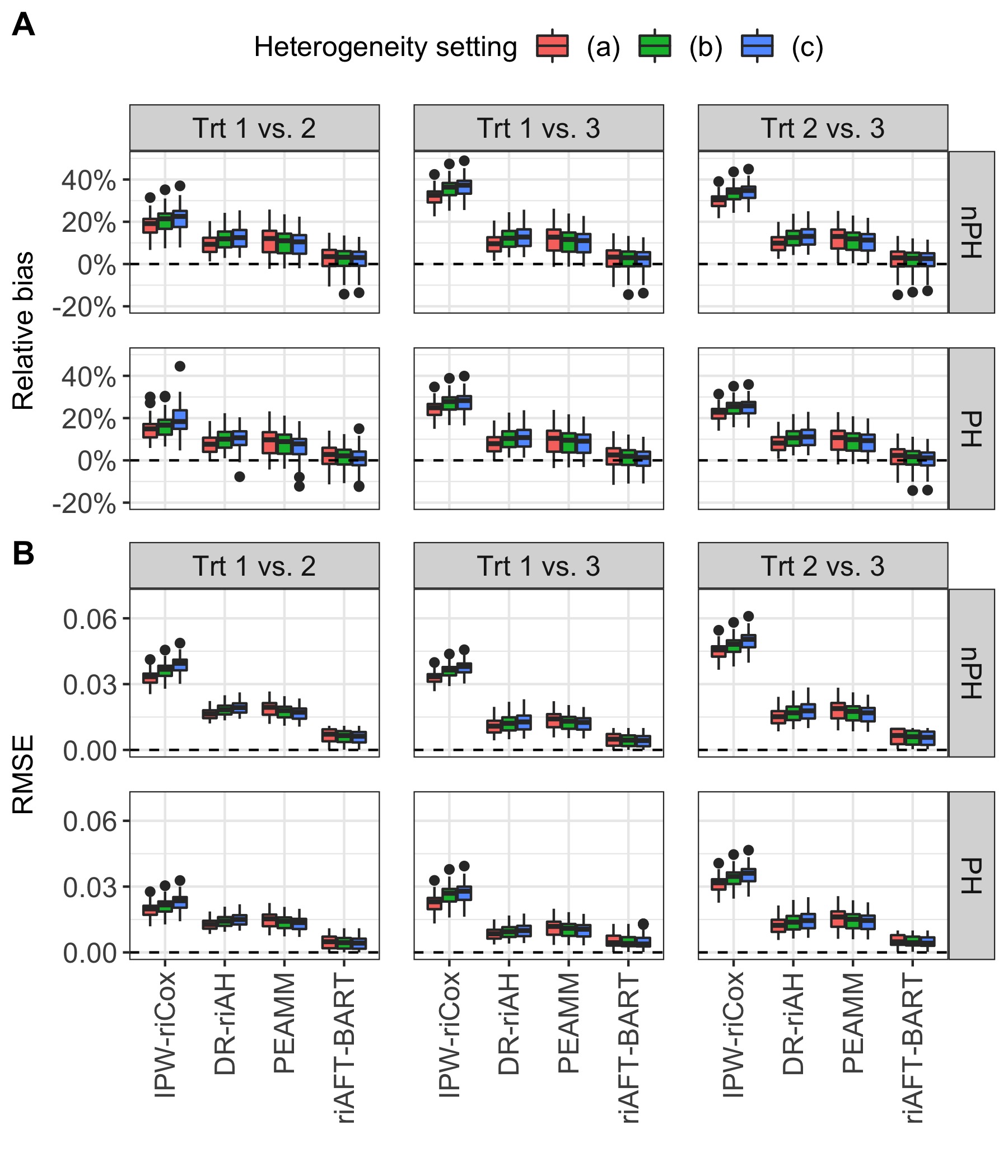}
    \caption{Relative biases (Panel A) and root-mean-squared-errors (RMSE) (Panel B) among 40 generalized propensity score subgroups under 6 data configurations: (heterogeneity settings a, b, c) × (proportional hazards (PH) and nonproportional hazards (nPH)) for each of four methods, IPW-riCox, DR-riAH, PEAMM and riAFT-BART. Three pairwise treatment effects were estimated by averaging the individual survival treatment effect (based on 3-week survival probability) across individuals in each subgroup.  Each boxplot visualizes the distribution of relative biases or the distribution of RMSE for 40 subgroups, each averaged across 250 simulation runs.}
    \label{fig:survprob_hs}
\end{figure}

Figure~\ref{fig:survprob_hs} displays the relative biases and RMSE results among 40 generalized propensity score subgroups, each averaged across 250 simulation runs. Three pairwise treatment effects were estimated by averaging the ISTE $\hat{\zeta}_{a_j,a_{j'}}(\bm{x}_{ik})$ (based on 3-week survival probability) across individuals in each subgroup. Under both settings of proportional hazards and nonproportional hazards, the proposed method riAFT-BART boasts the smallest biases and RMSE across all three treatment effects, particularly for subpopulation near the centroid of the generalized propensity score distribution. In sharp contrast, IPW-riCox, which encodes parametric and linearity assumptions and assumes proportional hazards, yielded the largest biases and RMSE. The increasing complexity level of treatment effect heterogeneity did not impact the performance of  more flexible methods PEAMM and riAFT-BART, but induced larger biases and RMSE for parametric methods IPW-riCox and DR-riAH. The violation of the proportional hazards assumption, which introduced an additional data complexity (the rate parameter $\eta$ depends on the covariate $x_{ik1}$), had a negligible effect on the performance of riAFT-BART, but all three comparison methods had a deteriorated performance. It is worth noting that all methods produced large biases and RMSE for subpopulation at the tail regions of the generalized propensity score distribution, suggesting that causal inferences drawn about this subpopulation may not be reliable. Results based on 3-week RMST, shown in Web Figure 3, convey the same messages.

\begin{table}[!htb]
\centering
\caption{Mean and (standard deviation) of precision in the estimation of heterogeneity effects (PEHE) across 250 data replications for each of the 4 methods based on 3-week survival probability under six configurations: (proportional hazards [PH] vs. nonproporitonal hazards [nPH]) $\times$ (heterogeneity setting [HS] (a) vs. (b) vs. (c)).} 
\begin{tabular}{clccccccc}
\toprule
& &  \multicolumn{3}{c}{PH} && \multicolumn{3}{c}{nPH}  \\
\cmidrule{3-5}  \cmidrule{7-9}
& Methods &  HS(a) & HS(b) & HS(c) & & HS(a) & HS(b) & HS(c) \\
\midrule
\multirow{4}{*}{Trt 1 vs. 2}  & IPW-riCox & .061 (.028)	&	.067 (.029) &  .070 (.030) & & .082 (.028)	&	.089 (.029) &  .092 (.030)\\
& DR-riAH & .018 (.021)	& .020 (.022) & .026 (.023) && .028 (.020)	&	.033 (.021) & .035 (.021) \\
&PEAMM & .028 (.023)	& .024 (.022)	 & .020 (.021) && .038 (.024)	&   .034 (.023)	 & .031 (.023) \\
&riAFT-BART & 	.012 (.010) & .009 (.009) & .006 (.009)   && .017 (.012)  & .014 (.011)& .011 (.011) \\
\midrule
\multirow{4}{*}{Trt 1 vs. 3}  & IPW-riCox & .065 (.027)	&	.070 (.028) &  .073 (.031) & & .083 (.029)	&	.090 (.030) &  .093 (.031) \\
& DR-riAH & .021 (.018) 	& .026 (.020)  & .029 (.022) && .031 (.019)	&	.036 (.020) & .039 (.022) \\
&PEAMM & .035 (.024)	& .030 (.023)	 & .026 (.022) && .049 (.025)	&   .044 (.024)	 &  .037 (.023) \\
&riAFT-BART & 	.016 (.010) & .013 (.009) & .010 (.009)   && .022 (.014)  & .019 (.013)& .015 (.012)\\
\midrule
\multirow{4}{*}{Trt 2 vs. 3}  & IPW-riCox & .088 (.030)	&	.095 (.031) &  .098 (.032) & & .105 (.031)	&	.115 (.032) &  .118 (.033) \\
& DR-riAH & .038 (.023)	& 043 (.023) & .045 (.024) && .045 (.021)	&	.050 (.022) & .052 (.022)\\
&PEAMM & .049 (.023)	& .045 (.022) & .041 (.022) && .058 (.024)	&   .054 (.023)	 & .049 (.023)\\
&riAFT-BART & 	.017 (.012)& .015 (.011) & .013 (.011)   && .022 (.015) & .020 (.014)& .018 (.014)\\
\bottomrule
\end{tabular}
\label{tab:PEHE-survival-moderate}
\end{table}

\subsection{Variable selection} \label{sec:vs-sim}
We adopt the simulation settings used in  \cite{hu2021variable} to evaluate the performance of variable selection methods among incomplete, clustered survival data.  We used the multivariate amputation approach to generate missing data scenarios with desired  missingness percentages under the missing at random mechanism \citep{schouten2018generating}. The multivariate amputation procedure comprises three main steps. Firstly, the complete data is randomly divided into a specified number of subsets, each allowing for the specification of any missing data pattern. Secondly, weighted sum scores are calculated for individuals in each subset in order to amputate the data. These scores are determined based on the relationship between the variable to be amputated and the variables on which the missingness in the amputated variable depends. Finally, a logistic distribution function, as described by \cite{van2018flexible}, is applied to the weighted sum scores to compute the probability of missingness. This probability is then used to determine whether each data point becomes missing or not.
We investigated the practical operating characteristics of our proposed variable selection method using riAFT-BART, and compare it to PEAMM,  frailty models with penalized likelihood (FrailtyPenal), frailty models with penalized profile h-likelihood (FrailtyHL) and backward stepwise selection with random-intercept Cox regression model (riCox). Four performance metrics were used to evaluate the variable selection results: precision, recall, $F_1$ and Type I error (Section~\ref{sec:vs-mis}). 

The multivariate amputation procedure  was implemented via the $\code{ampute}$ function of the $\R$ package $\code{mice}$ \citep{schouten2018generating}.  Imputation was performed using the $\R$ package $\code{mice}$ \citep{buuren2010mice}. 
To implement our proposed method using riAFT-BART, we used 4500 posterior draws with the first 1000 discarded as burn-in. The permutation distributions of VIP scores were constructed from 100 riAFT-BART model runs. To implement the variable selection procedure based on PEAMM, we used the helper function (\code{as\_ped}) from $\R$ package \code{pammtools} to reformat the data and used the \code{gam} function from $\R$ package \code{mgcv} for variable selection. Variable selection by frailty models was implemented using the \code{frailtyPenal} function of $\R$ package \code{frailtypack} for penalized marginal likelihood; and using the function \code{frailty.vs} from $\R$ package \code{frailtyHL} for penalized profile h-likelihood. 
For backward stepwise selection, the random-intercept Cox regression model was fitted using the \code{coxme} function from $\R$ package \code{coxme}. 

\subsubsection{Simulation design}
The simulation scenarios are motivated by the data structures observed in the COVID-19 data set used in our case study. 
We considered $K=10$ clusters, each with a sample size of $n_k=200$. We generated 8 \emph{useful} predictors that are truly related to the survival outcomes, $X_{ik1}, \ldots, X_{ik8}$, and 20 noise predictors, $X_{ik9}, \ldots, X_{ik28}$. 
We simulated $X_{ik1}$ and $X_{ik2}$ independently from  $\text{Bern}(0.5)$, $X_{ik3}$, $X_{ik4}$ from the standard normal distribution $N(0,1)$, and $X_{ik5}, X_{ik6}, X_{ik7}, X_{ik8}$ were designed to have missing values under the missing at random mechanism. For a missing at random predictor variable, we specify the true forms in which  the predictor depends on the other predictors as the following: 
\begin{eqnarray*}
&x_{ik5} \cond x_{ik2},x_{ik3} \sim N(0.3x_{ik2}-0.2x_{ik3},1)\\
&x_{ik6} \cond x_{ik3},x_{ik4} \sim N(-0.4x_{ik3}+0.4x_{ik4}+0.3x_{ik3}x_{ik4},1)\\
&x_{ik7} \cond x_{ik4},x_{ik5},x_{ik6} \sim N(0.1x_{ik4}(x_{ik5}-2)^2-0.1x_{ik6}^2,1)\\
&x_{ik8} \cond x_{ik5},x_{ik6},x_{ik7} \sim N(-0.3x_{ik5}^2+0.5x_{ik6}+0.3x_{ik7}+0.2x_{ik6}x_{ik7},1).
\end{eqnarray*}
Among the 20 noise predictors, 10 were generated as continuous variables $X_{ik9}, \ldots, X_{ik18} \stackrel{i.i.d}{\sim}N(0,1)$,  and the other 10 were generated as binary variables $X_{ik19}, \ldots, X_{ik28} \stackrel{i.i.d}{\sim}\text{Bern}(0.5)$.

Similar to the outcome generating process described in Section~\ref{sec:sim-design-teh}, we generate the observed and uncensored survival times from the Weibull survival distribution, 
\begin{eqnarray*}\label{eq:Tmod}
 T_{ik} = \lsq \frac{-\log U}{\lambda\exp\lbc q(\bm{X}_{ik},b_k) \rbc} \rsq ^{1/\eta}
\end{eqnarray*}
where $\lambda=3000$, $U$ is a random variable following the uniform distribution on the interval $[0,1]$, $b_k\sim N(0,4^2)$, and
\begin{eqnarray}\label{eq:vs-qmod}
q(\bm{X}_{ik},b_k)&=&1.8x_{ik1}+0.5x_{ik2}+1.1x_{ik3}-0.4e^{x_{ik5}}+0.4(x_{ik6}-1.5)^2 \\\nonumber
&&+0.1(x_{ik7}-0.1)^3-5\sin(.314 x_{ik4}x_{ik8}) -0.4x_{ik5}x_{ik7}+b_k
\end{eqnarray}
We further generated the censoring time $C$ independently from an exponential distribution with the rate parameter selected to induce the censoring proportion of 50\%. The parameter $\eta$ was  set to 2 and $\exp{(0.7+0.5x_{ik1})}$ to respectively produce proportional hazards and nonproportional hazards. This data generating process was designed to make it difficult for any method to accurately model the survival outcome. We considered the outcome model with arbitrary data complexity that reflects common situations in health datasets: (i) discrete predictors with strong ($X_{ik1}$) and moderate ($X_{ik2}$) associations; (ii) both linear and nonlinear forms of continuous predictors ($X_{ik6}$ and $X_{ik7}$); (iii) nonadditive effects ($X_{ik4}X_{ik8}$).  

After generating the fully observed data, we amputated the four predictors designed to have missing values $X_{ik5}, X_{ik6}, X_{ik7}, X_{ik8}$  under the missing at random mechanism using the multivariate amputation approach. Additional technical details for the amputation procedure are available in \cite{schouten2018generating}; and additional amputation details for the simulation appear in Web Section 2. The resultant simulation data have an overall missingness proportion of 40\%. The process of data generation was replicated 250 times.

\subsubsection{Simulation results}
Table~\ref{tab:sim-res-vs} summarizes variable selection results by each method. 
The optimal performance in the presence of incomplete data is demonstrated by selecting the most suitable threshold value for $\pi$, which is determined based on the $F_1$ score (refer to Web Figures 4-5). This performance is then compared with that of fully observed data (prior to amputation) and complete cases exclusively. It is noteworthy that the optimal threshold value of $\pi$ for riAFT-BART is 0.1, corresponding to the highest $F_1$ score. Additionally, the $F_1$ score declines at elevated values of $\pi$, attributable to reduced recall values.

The proposed method riAFT-BART has the best performance across all three data panels (fully observed, with missing data, complete cases), evidenced by the highest precision, recall and $F_1$ score, and the lowest type I error. The second best performing method is PEAMM, for its modeling flexibility over parametric methods. In the presence of missing covariate data, all methods had a slightly deteriorated performance when performing variable selection in combination with bootstrap imputation, but the performance of variable selection  among only complete cases was considerably lower.

\setlength{\tabcolsep}{3pt} 
\begin{table}[!htb]
    \centering
    \caption{Simulation results for each variable selection approach performed on the fully observed data, and among incomplete data with 40\% overall missingness in covariates under proportional hazards and non-proportional hazards. Imputation was conducted via $\code{mice}$. For each of five methods, we show results corresponding to the best threshold values of $\pi$ (based on $F_1$).} 
    \bgroup
\def\arraystretch{1.1} 
    \begin{tabular}{lccccclcccc}
    \toprule
         & \multicolumn{4}{c}{Proportional hazards} &&& \multicolumn{4}{c}{Non-proportional hazards} \\\cline{2-5}\cline{8-11}
         &Precision &Recall &$F_1$  &Type I error &&&Precision &Recall &$F_1$  &Type I error\\\hline 
      \multicolumn{11}{c}{\textbf{Fully observed data}}\\
         riAFT-BART & 0.95&0.88&0.91&0.01&&riAFT-BART&0.93&0.87&0.89&0.02\\
        PEAMM & 0.83&0.82&0.82&0.02&&PEAMM&0.81&0.80&0.80&0.03\\
          FrailtyPenal & 0.79 &0.76&0.77&0.04&&FrailtyPenal&0.75&0.69&0.72&0.05\\
        FrailtyHL & 0.81 &0.78&0.79&0.03&&FrailtyHL&0.78&0.71&0.74&0.04\\
        riCox & 0.78&0.72&0.75&0.05&&riCox&0.72&0.67&0.69&0.05\\
         \midrule
    \multicolumn{11}{c}{\textbf{40\% overall missing}}\\
        riAFT-BART $\pi=0.1$  & 0.93&0.85&0.89&0.02&&riAFT-BART $\pi=0.1$ &0.89&0.84&0.86&0.03\\
        PEAMM  $\pi=0.7$ & 0.88&0.74&0.80&0.03&&PEAMM $\pi=0.6$&0.84&0.72&0.78&0.04\\
        FrailtyPenal  $\pi=0.6$ & 0.76&0.71&0.74&0.04&&FrailtyPenal $\pi=0.7$&0.72&0.67&0.69&0.05\\
        FrailtyHL  $\pi=0.6$ & 0.79&0.73&0.76&0.05&&FrailtyHL $\pi=0.7$&0.75&0.70&0.71&0.04\\
          riCox  $\pi=0.8$ & 0.85&0.67&0.73&0.04&&riCox $\pi=0.7$&0.71&0.63&0.68&0.07\\
         \midrule
        \multicolumn{11}{c}{\textbf{Complete cases}}\\
        riAFT-BART & 0.85&0.81&0.83&0.03&&riAFT-BART&0.83&0.79&0.81&0.04\\
        PEAMM & 0.77&0.73&0.75&0.02&&PEAMM&0.73&0.71&0.72&0.03\\
       FrailtyPenal & 0.71&0.65&0.69&0.05&&FrailtyPenal&0.69&0.62&0.65&0.04\\
        FrailtyHL & 0.73&0.68&0.71&0.04&&FrailtyHL&0.70&0.65&0.67&0.05\\
         riCox & 0.71&0.66&0.68&0.04&&riCox&0.66&0.62&0.64&0.03\\
    \bottomrule
    \end{tabular}
  \egroup  
    \label{tab:sim-res-vs}
\end{table}

\begin{figure}[!ht]
    \centering
    \includegraphics[width=1\textwidth]{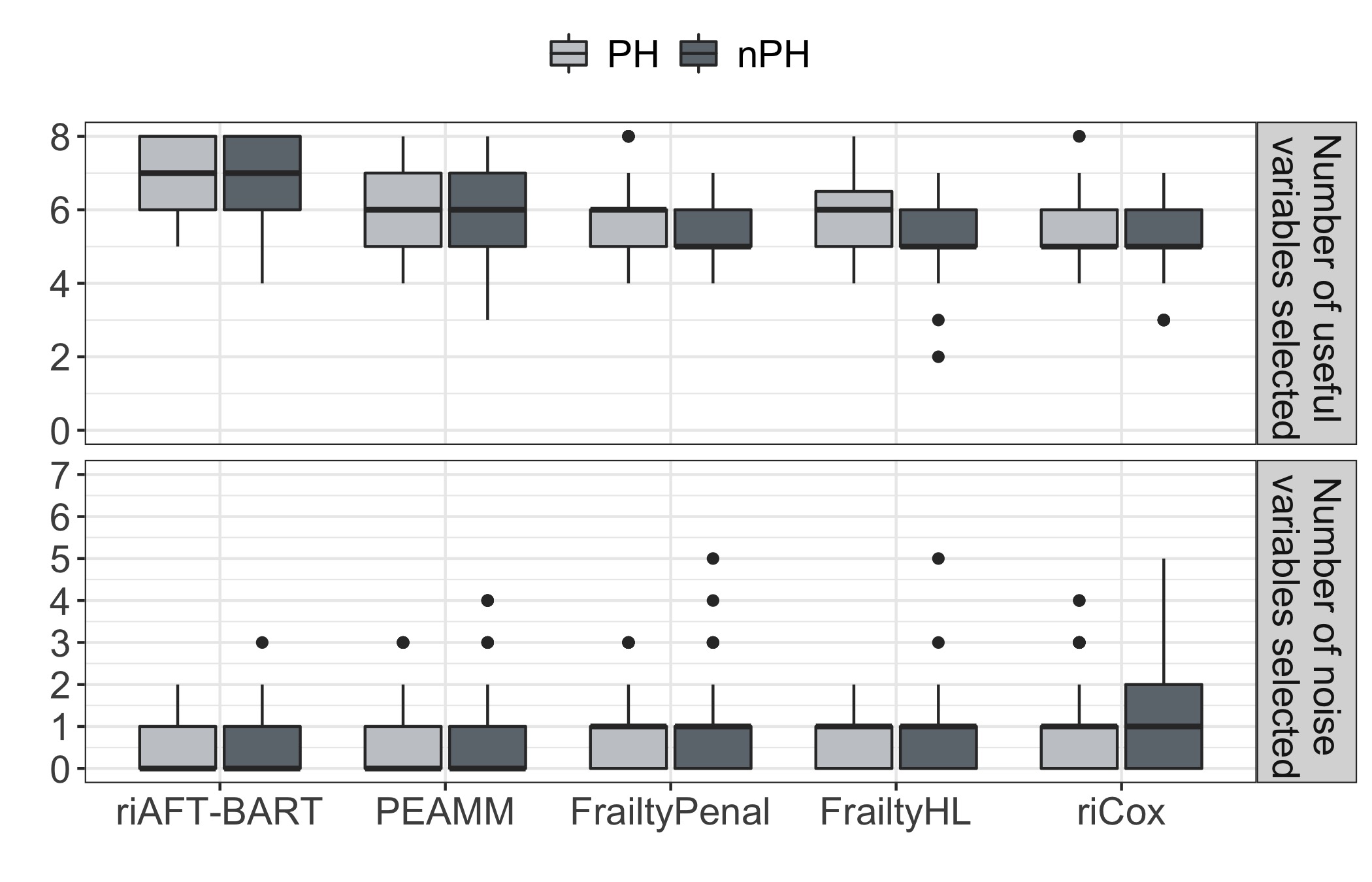}
    \caption{The distribution, across 250 data replications, of the numbers of selected noise predictors and useful predictors for each of five methods: riAFT-BART, PEAMM,  FrailtyHL, FrailtyPenal and riCox,  with clustered survival data generated under both proportional hazards (PH) and non-proportional hazards (nPH). The total number of useful predictors is 8 and the total number of noise predictors is 20. There are $K=10$ clusters, each with a size of 200; the total sample size is 2000. The overall proportion of missingness is 40\%.} 
    \label{fig:n_variables_selected}
\end{figure}

Demonstrated by Figure~\ref{fig:n_variables_selected}, the proposed method riAFT-BART tended to select the most useful predictors and the least noise predictors, followed by another flexible method PEAMM, under both proportional hazards and non-proportional hazards. Among parametric methods, penalized likelihood methods for frailty models had slightly better performance than riCox. All methods tended to select more noise predictors and less useful predictors under non-proportional hazards; the decrease in the performance due to elevated data complexity associated with nonproportional hazards, was only marginal for riAFT-BART but considerable for less flexible methods, particularly FrailtyHL and riCox.

\begin{figure}[!ht]
    \centering
    \includegraphics[width=1\textwidth]{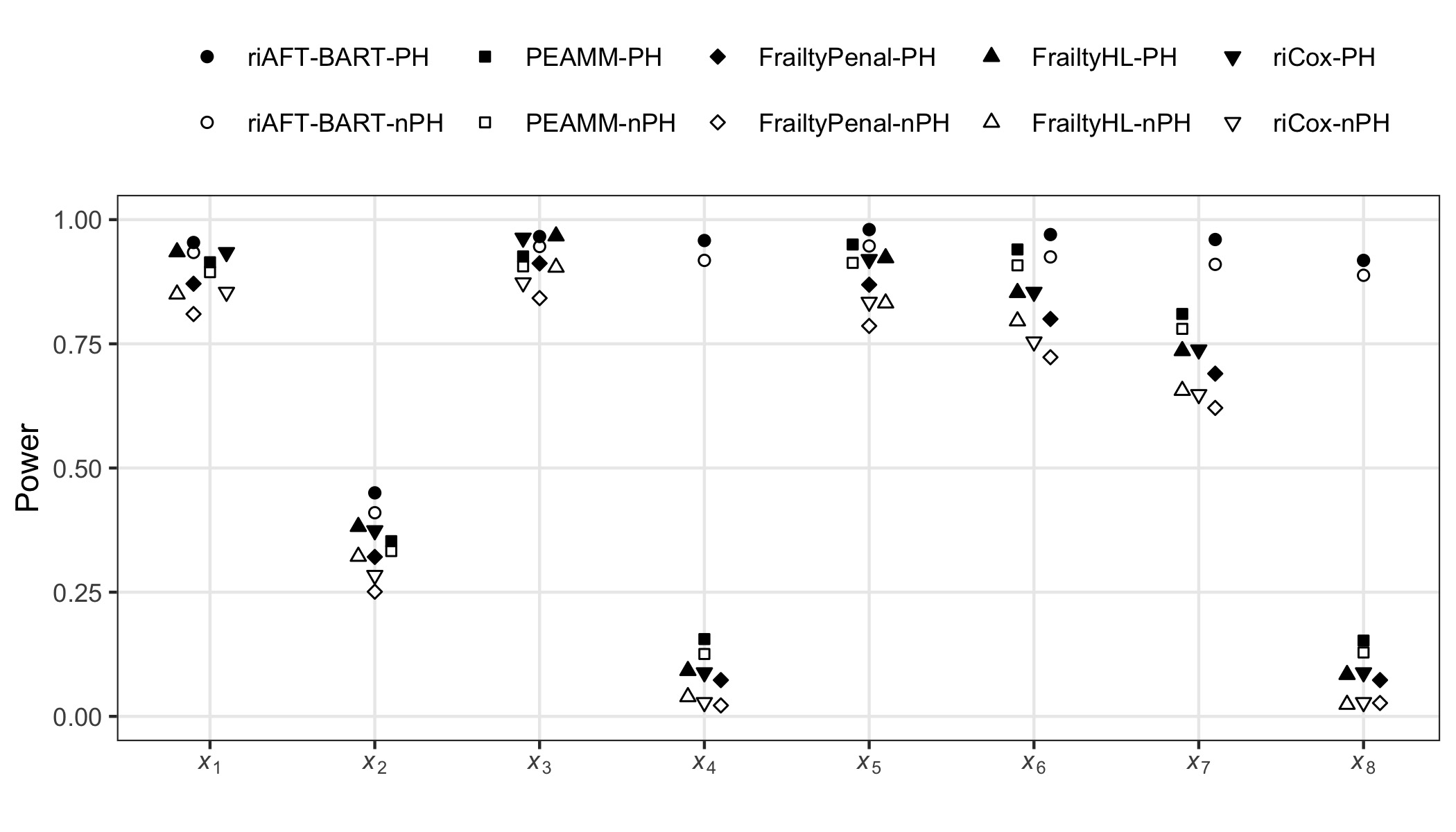}
    \caption{Power of each of five methods: riAFT-BART, PEAMM,  FrailtyHL, FrailtyPenal and riCox, for selecting each of 8 useful predictors with clustered survival data generated under proportional hazards (PH) and non-proportional hazards (nPH), based on 250 data replications. There are $K=10$ clusters, each with a size of 200; the total sample size is 2000. The overall proportion of missingness is 40\%. Filled symbols represent the PH setting, and open symbols correspond to the nPH setting. } 
    \label{fig:vs-power}
\end{figure}

A perusal of each method's ability to select each of useful predictors is visualized in Figure~\ref{fig:vs-power}. All methods had a good power ($>0.8$) for selecting a strong discrete predictor ($X_{ik1}$), but unsatisfactory power ($<0.5$) for detecting a discrete predictor ($X_{ik2}$) that only has a moderate relationship with the outcome. For continuous variables, all methods performed well in selecting $X_{ik3}$, which has only a linear relationship with the log survival time, but riAFT-BART produced a significantly higher power for selecting the predictor $X_{ik4}$, which has a complex nonadditive form in model~\eqref{eq:vs-qmod}. Among the four predictors with missing data $X_{ik5}, \ldots, X_{ik8} $, the advantage of riAFT-BART is demonstrated by the higher power of selecting predictors that have more complex functional forms that are difficult to be captured by other methods considered.  A direct comparison of the proportional hazards setting with nonproportional hazards shows that all methods performed better under proportional hazards.  However, when the proportional hazards assumption is violated, the effects are least detrimental for PEAMM and riAFT-BART.

As our proposed method riAFT-BART pivots on the predictor's VIP, we examined the convergence of the Markov chain by plotting 4500 posterior  draws, with the first 1000 discarded as burn-in, of the VIP for four useful predictors and four noise predictors in Web Figure 6. The method converged well.    

All simulation procedures were conducted in $R$ on an iMac equipped with a 4 GHz Intel Core i7 processor. The dataset under consideration was comprised of 10 clusters ($K=10$), each with a sample size of 200 ($n_k=200$), and included 8 relevant predictors and 20 noise predictors. In terms of processing time, riAFT-BART required approximately 58 minutes to execute, PEAMM took around 42 minutes, FrailtyPental ran for about 28 minutes, FrailtyHL consumed roughly 30 minutes, and riCox finished in about 8 minutes. These methods were run with the default memory size setting of 100MB.

\section{Case study: Predictors for in-hospital mortality and heterogeneous survival treatment effects among post-ICU COVID patients}
\label{sec:application}
In this case study, we first applied the proposed variable selection methods to select important predictors for in-hospital mortality using  a comprehensive COVID-19 dataset, and then evaluated the treatment effect heterogeneity on patient survival using the selected predictors by implementing our proposed method riAFT-BART. The COVID-19 dataset 
was drawn from six hospitals (clusters) of the Epic electronic medical records system at the Mount Sinai Medical Center. Included in this dataset were 1955 patients who were diagnosed with COVID-19 between March 10, 2020 to February 26, 2021 and were in intensive care (ICU) due to COVID-19 during their hospital stay \citep{wang2020hospitalised}. Post ICU admission, each patient received COVID-specific medications: either dexamethasone, or remdesivir, or the combination of dexamethasone and remdesivir (dexamethasone+remdesivir). The cluster sizes and number of patients in each treatment group are provided in Web Table 4. 

We used the following 28 baseline (time of ICU admission) variables: age, sex, self-reported race and ethnicity, smoking status,  binary comorbidities including hypertension, coronary artery disease, cancer, diabetes, asthma, and chronic obstructive pulmonary disease,  vital signs  including temperature, systolic blood pressure, diastolic blood pressure, patient oxygen level (definition in Web Table 3), heart rate, the fraction of inspired oxygen, body mass index, oxygen saturation, risk score (sequential organ failure assessment score and glasgow coma scale), laboratory test results including partial pressure of oxygen, D dimer value and inflammatory markers such as lactate dehydrogenase, ferritin, c-reactive protein, creatinine and white blood cell count.  Vital signs, risk score and lab results were measured during the first 24 hours of ICU admission. 
Among the 28 candidate predictor variables, 17 variables have missing data and the missingness proportion ranges from 1.4\% to 28.6\%. Only 1187 (60.7\%) patients have fully observed data for all covariates, which motivated the overall missingness proportion of 40\% used in our simulation study (Section~\ref{sec:vs-sim}). Detailed summary statistics for baseline variables appear in Web Table 4. 

The Kaplan-Meier survival curves by treatment group are displayed in Web Figure 7, with crossovers suggesting nonproportional hazards. Assessed by the distribution of the generalized propensity scores  shown in Web Figure 8, there is moderate covariate overlap across three treatment groups in the COVID-19 data. 

\begin{table}[htbp]
    \centering
    \caption{Variable selection results by each of 5 methods, with the best selection threshold value of $\pi$ suggested in simulations under non-proportional hazards. riAFT-BART and riCox both selected 8 variables, PEAMM selected 7 variables, FrailtyPenal selected 6 variables and FrailtyHL selected 9 variables. } 
    \begin{tabular}{lccccc}
    \toprule
        Variables & riAFT-BART & PEAMM & FrailtyPenal  & FrailtyHL &riCox  \\
        \midrule
         Oxygen level & Yes & Yes & No & Yes & Yes \\
         Age & Yes & Yes & Yes & Yes & No \\
         Creatinine & Yes & Yes & No & Yes & Yes\\
         White blood cell & Yes & Yes & No & No & Yes\\
         SOFA score & Yes & Yes & Yes & Yes & No\\
         Race & Yes & No & Yes & Yes & Yes\\
         D dimer & Yes & No & No & Yes & No\\
         LDH & Yes & Yes & Yes & Yes & Yes\\
         C-reactive protein & No & Yes & No & Yes & Yes \\
         Ferritin & No & No & No & Yes & Yes \\
         Oxygen saturation & No & No & Yes & No & No \\
         Glasgow coma scale & No & No & Yes & No & Yes \\
         \bottomrule
    \end{tabular}
\footnotesize \\\smallskip
Abbreviations:  LDH = Lactate dehydrogenase; SOFA = Sequential organ failure assessment
    \label{tab:covid-sel-vbls}
\end{table}

\begin{figure}[!ht]
\centering
\includegraphics[width=0.8\textwidth]{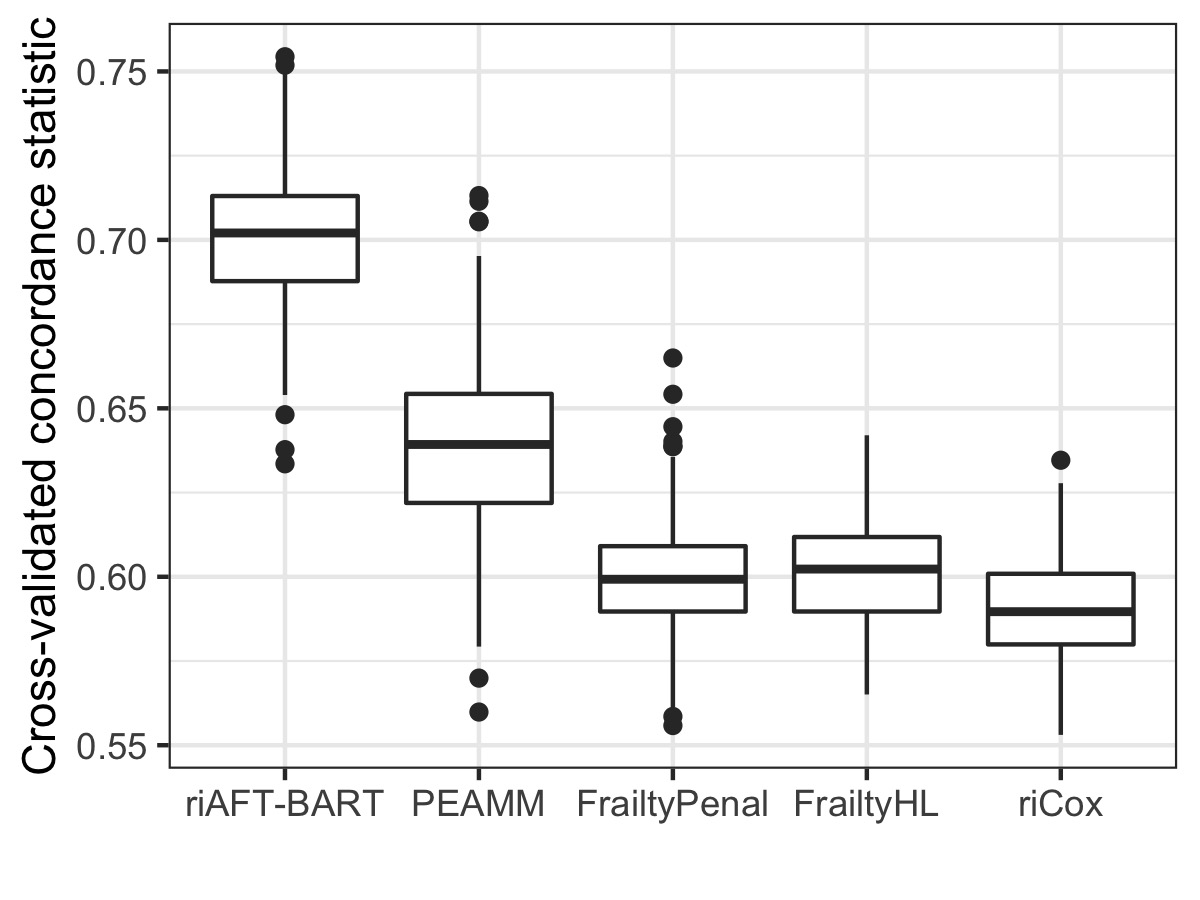}
\caption{The distribution of cross-validated concordance statistics across 250 data replications for each of five methods using the COVID-19 dataset.} 
\label{fig:cv-concordance}
\end{figure}

We first implemented all five variable selection methods.  All 28 candidate predictors were included in the imputation models, and the optimal threshold value of $\pi$ for each method, which produced the best $F_1$ score in the simulation study, was used to select the final set of predictors. As summarized in Table~\ref{tab:covid-sel-vbls}, FrailtyHL selected the most (9) predictors,  riAFT-BART and riCox both selected 8 predictors, PEAMM selected 7 preditors and FrailtyPenal selected the least (6) predictors.  Age and lactate dehydrogenase were the only two predictors selected by all five methods. Variables selected by the two flexible methods riAFT-BART and PEAMM were largely in agreement with each other, with the exception that riAFT-BART but not PEAMM selected Race and D dimer, and PEAMM but not riAFT-BART selected C-reactive protein.

To further distinguish between methods based on their ability to select the most relevant predictors, we computed the predictive performance of the five methods that modeled the outcome variable on their respectively selected predictor variables. For survival outcomes, we used the  concordance statistic as a metric to assess the predictive performance \citep{harrell1996multivariable}.  The concordance statistic is the  fraction of concordant pairs among all pairs of individuals. A pair of patients is called concordant if the patient with higher (lower) predicted survival probability at a given time point had a longer (shorter) time to death.  In this case study, we chose 21 days post ICU admission as the target time point, because it is a sufficiently long time to evaluate the effects of COVID-19 treatments on in-hospital mortality for severe patients.  As the COVID-19 dataset is subject to missing covariate data, to calculate the cross validated concordance statistic, we
follow the strategy described in \cite{lin2022a}. We 
first split the data into two halves and implemented each of the five variable selection methods on half of the data. Then we imputed the other half of the data with a single imputation and recorded the concordance statistic. Finally we repeated the previous two steps 250 times to get the distribution of the cross-validated concordance statistics. Figure~\ref{fig:cv-concordance} shows that the proposed method riAFT-BART yielded the highest predictive performance with a median concordance statistic $>0.7$.  

Supported by both the simulation study and case study, riAFT-BART provided the best variable selection results. Thus we used the 8 predictors selected by riAFT-BART as confounding variables, and estimated the heterogeneous treatment effects on patient survival using methodology described in Section~\ref{sec:TEH}. To address the missing data issue when drawing causal inferences about treatment effects, we fitted an riAFT-BART model~\eqref{eq: riAFT-BART} on each of the 100 bootstrap and imputed datasets (previously conducted for variable selection), and drew statistical inferences about the ISTE $\zeta_{a_j,a_{j'}}(\bm{x}_{ik})$, defined in equation~\eqref{eq:ISTE}, by pooling posterior samples of the ISTE, $\tilde{f}^d(a_j,\bm{x}_{ik}) - \tilde{f}^d(a_{j'},\bm{x}_{ik})$ as in equation~\eqref{eq:theta_hat}, across model fits arising from the multiple datasets \citep{zhou2010note,hu2022aflexible}. Inferences about the population average treatment effects $\theta_{a_j,a_{j'}}$ can be obtained from the posterior samples of  $\tilde{f}^d(a_j,\bm{x}_{ik}) - \tilde{f}^d(a_{j'},\bm{x}_{ik})$ across the sample population \citep{hu2022flexible}. The population average treatment effects, summarized in Web Table 5, suggest that there was no statistically significant difference in treatment effect when comparing dexamethasone with remdesivir or with a combined use of dexamethasone and remdesivir. However, there was a statistically significant treatment benefit associated with dexamethasone+remdesivir compared to using remdesivir alone. Web Figure 9 demonstrates that substantial variability in the institutional effect was captured by the riAFT-BART model: the Mount Sinai main hospital had considerably better outcomes than the Mount Sinai Queens hospital.

\begin{figure}[!ht]
    \centering
    \includegraphics[width=1\textwidth]{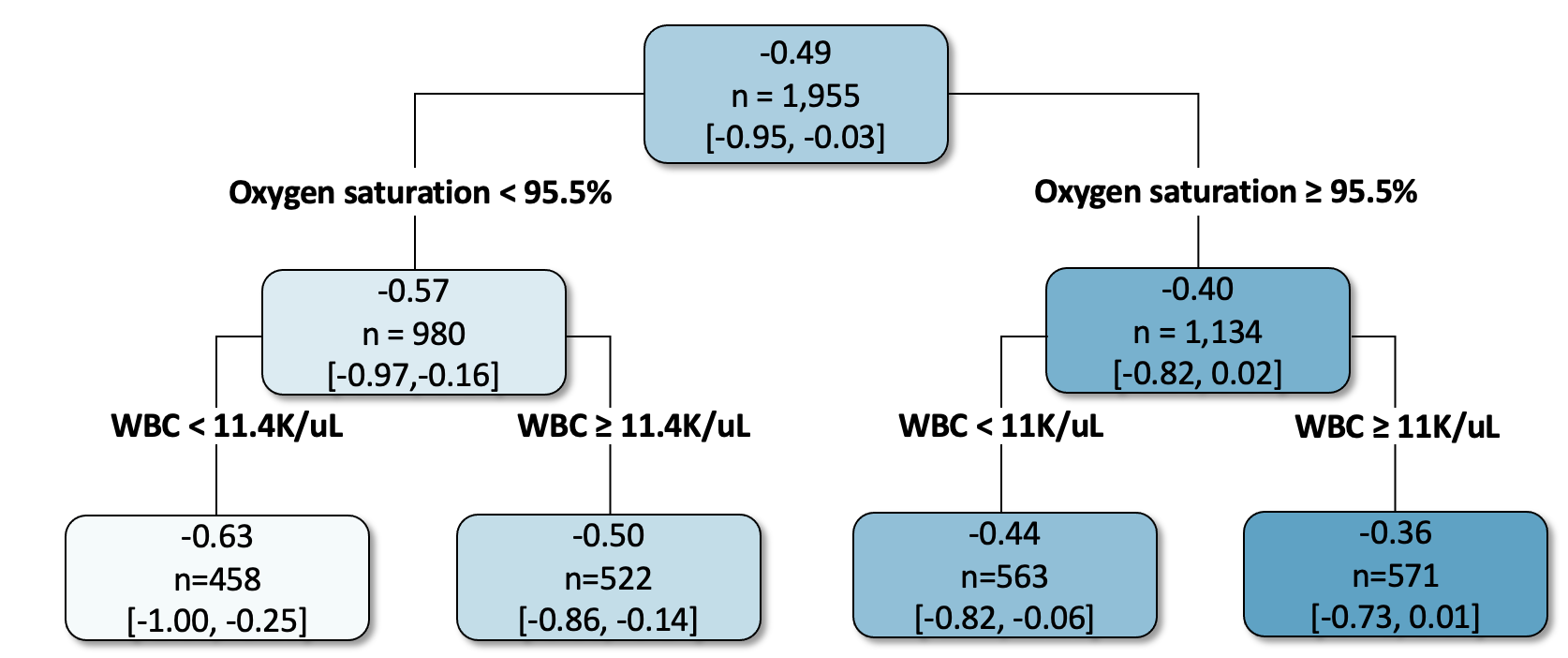}
    \caption{Final Random Forests model fit to the posterior mean of the individual survival treatment effect comparing remdesivir and dexamethasone + remdesivir. Values in each node correspond to the posterior mean, in terms of difference in log survival days, for the subgroup of individuals represented in that node. Uncertainty intervals were obtained by pooling the posterior samples arising from the multiple imputed data sets. WBC: White blood cell.} 
    \label{fig:covid-teh}
\end{figure}

Turning to the treatment effect heterogeneity. Figure~\ref{fig:covid-teh} summarizes the combination rules of covariates $\mathcal{S}_{a_j,a_{j'}}$ that have differential subgroup treatment effect $\mathcal{M}(\mathcal{S}_{a_j,a_{j'}})$, defined in equation~\eqref{eq:MS-subg}, than the sample population treatment effect. 
When comparing remdesivir and dexamethasone + remdesivir across the sample population, there was significant treatment benefit, suggested by a longer expected survival time, for the combined use of dexamethasone and remdesivir. Our analysis suggests that combining dexamethasone and remdesivir offered enhanced treatment benefit for relatively healthier COVID-19 patients with lower oxygen saturation  ($<95.5\%$) and a normal white blood cell count ($<11.4 K/\mu L$), demonstrated by the leftmost branch of the tree diagram in Figure~\ref{fig:covid-teh}. By contrast, the rightmost branch suggests that there was no statistically significant treatment benefit associated with either treatment choice for  comparatively unhealthier patients with a higher level of oxygen saturation ($>95.5\%$). 


\section{Discussion} \label{sec:discussion}
We describe two important and practical utilities of a new Bayesian machine learning method riAFT-BART for analyzing clustered and censored survival data. To  address the implications of complex data structures of large-scale clinical datasets generated from multiple institutions for causal inferences about population treatment effect, riAFT-BART was developed to accurately estimate the population average treatment effect on patient survival while accounting for the variation in institutional effects \citep{hu2022flexible}.

Moving from broad applicability of a population study to personalized medicine requires the understanding of treatment effect heterogeneity. The treatment effect heterogeneity was traditionally identified by first enumerating potential effect modifiers with subject-matter experts and then estimating the average treatment effect within each subgroup \citep{kent2020predictive}. This approach is particularly suitable to confirmatory treatment effect heterogeneity analysis in randomized clinical trials. In observational data with a large number of pre-treatment covariates, such \emph{a priori} specification that separates the issues of confounding and treatment effect heterogeneity is often practically infeasible. We described a way in which the new method riAFT-BART can be used to perform an exploratory treatment effect heterogeneity analysis to generate  scientifically meaningful hypotheses and treatment effect discovery. Through a comprehensive simulation representative of complex confounding and heterogeneity settings with right-censored survival data, we provided new empirical evidence for the better performance of our proposed method in comparison with three existing methods.  In the application to a large COVID-19 dataset, we found that oxygen saturation and white blood cell count were  two key patient factors that modulated the comparative treatment
effect between remdesivir and dexamethasone + remdesivir. We further exploited the posterior samples of the ISTE  from riAFT-BART and the random forests model to explore clinically meaningful relationships between COVID-19 medications, patient profile and expected survival time. 
The results could facilitate treatment effect discovery in subpopulations and may inform personalized treatment strategy and the planning of future confirmatory randomized trials. Based on the simulation results, caution is warranted when drawing causal inferences about subpopulations in the tail regions of the generalized propensity score distribution, as the reliability of these inferences may be compromised. 

Drawing on prior work \citep{hu2021variable} that combines bootstrap imputation and BART for variable selection among incomplete data, we develop a way in which riAFT-BART can be used to select important predictors among incomplete, clustered and censored survival observations. This strategy is general enough to accommodate any missing data pattern and is highly flexible as it leverages BART to model the relationship between survival times and covariates. As demonstrated by both simulations and the case study, this elevated modeling flexibility gave rise to substantially better variable selection results, particularly in terms of identifying predictors of complex functional forms, compared to existing variable selection methods suitable for clustered survival data. In general, we recommend using smaller values for the selection threshold, such as $\pi = 0.1$ or $\pi = 0.2$, when employing riAFT-BART for variable selection. Furthermore, the determination of the optimal selection threshold value can be achieved through cross-validation. Specifically, the most suitable threshold value ought to yield a set of selected variables that result in better cross-validated predictive performance.

There are several important avenues for further research. First, the  riAFT-BART model could be extended to include the random slopes and  
accommodate the cluster-level covariates. Correspondingly, a variable selection method can be developed to identify important predictors at both the individual level and the cluster level. Second,  using flexible modeling techniques alone will not address violations of the no unmeasured confounding assumption required for causal inference. Developing a formal sensitivity analysis approach \citep{hu2022aflexible} for unmeasured confounding would be a worthwhile and important contribution. Finally, our bootstrap imputation-based variable selection procedure is can be computationally demanding on a alrge dataset. Although the bootstrap resampling can be computed in parallel on multiple cores when such resources are available, it would be worthwhile to develop strategies that could offer substantial computational savings. 
\begin{acknowledgement}
This work was supported in part  by award ME-2021C2-23685 from the Patient-Centered Outcomes Research Institute, and by grants R21CA245855 and 1R01HL159077-01A1 from the National Institute of Health.
\end{acknowledgement}
\vspace*{1pc}

\noindent {\bf{Data Availability Statement}}

\noindent {\it{$\R$ codes to implement our proposed methods and the comparison methods, and to replicate our simulation studies are provided in the GitHub page of the author \url{https://github.com/liangyuanhu/TEH-VS-riAFTBART}, and are also available in the $\R$ package $\code{riAFTBART}$. Access to the COVID-19 data used in the case study needs to be requested and approved by the Icahn School of Medicine at Mount Sinai. The ``Data \& Code'' section in the supplemental materials provides instructions to reproduce all manuscript tables and figures using our $\R$ scripts. }}

\noindent {\bf{Conflict of Interest}}

\noindent {\it{The author has declared no conflict of interest. }}

\newpage
\bibliographystyle{biom} 
\bibliography{references} 

\end{document}


\maketitle

\section{Sampling algorithm of riAFT-BART}

With riAFT-BART, a Metropolis within Gibbs procedure was employed for posterior inferences about treatment effects on patient survival. 
The observed responses $y_{ik}$ were first centered via the following two steps: (i) fit a parametric intercept-only accelerated failure time model assuming log-normal residuals, and estimate the intercept $\hat{\mu}_{AFT}$ and the residual scale $\hat{\sigma}_{AFT}$; (ii) transform the responses as $y_{ik}^{cent} = y_{ik}\exp\lp-\hat{\mu}_{AFT} \rp$. 

We  use data augmentation to deal with right censoring \citep{henderson2020individualized}.  Working with the centered responses $y_{ik}^{cent}$, when $\Delta_{ik} =0$, we impute the unobserved and centered survival times $z_{ik}$ from a truncated normal distribution: 
$$\lsq \log Z_{ik} \cond \log Z_{ik}>\log y^{cent}_{ik} \rsq \sim N_{(\log y_{ik}^{cent},\infty)}\lp f(A_{ik}, \bm{X}_{ik})+b_k,\sigma^2\rp$$ 
in each Gibbs iteration, where  $\log Z_{ik} \sim N \lp f(A_{ik}, \bm{X}_{ik})+b_k, \sigma^2 \rp$.  The centered complete-data survival times are 
\[
  y^{cent,c}_{ik} = 
  \begin{cases}
    y^{cent}_{ik} \text{ \; \; if } \Delta_{ik} = 1 \\
    z_{ik} \text{ \; \; \;  if } \Delta_{ik} = 0 
  \end{cases}.
\]

Using the centered complete-data survival times $y_{ik}^{cent,c}$, the joint posterior is 
\begin{align*}
    &\hspace{12pt} P\left(b_k, \tau^2, \alpha_k, \mu_{lh}, \sigma^2 \cond y^{cent,c}_{ik}, \bm{X}_{ik}, A_{ik}, V_k,\{\mathcal{W}_h, \mathcal{M}_h\} \right)\\
    &\propto P \left(y^{cent,c}_{ik} \cond \bm{X}_{ik}, A_{ik}, V_k,b_k, \tau^2, \alpha_k, \sigma^2, \{\mathcal{W}_h, \mathcal{M}_h\} \right) P\left( b_k \cond \tau^2, \alpha_k \right) P\left( \tau^2 \right) P\left( \alpha_k \right) P\lp \mu_{lh}\rp P\left(\sigma^2\right), 
\end{align*}
where  $\mathcal{W}_h$ is the $h$th binary tree structure,  $\mathcal{M}_h = (\mu_{1h}, \ldots, \mu_{{c_h}h} )^T$ is the set of $c_h$ terminal node parameters associated with tree structure $\mathcal{W}_h$.  For a given value $A_{ik}$ and $\bm{X}_{ik}$ in the predictor space, the binary tree function returns the parameter $\mu_{lh}, l \in \{1, \ldots, c_h\}$ associated with the terminal node of the predictor subspace in which $\{A_{ik}, \bm{X}_{ik}\}$ falls.
 
We can draw the values of BART sum-of-trees model parameters, $\mu_{lh}$ and $\sigma^2$, directly from the fitted BART model. Their posterior distributions $P \lp \mu_{lh} \cond y^{cent,c}_{ik}, \bm{X}_{ik}, A_{ik}, V_k, b_k, \tau^2, \alpha_k, \sigma^2, \{\mathcal{W}_h\} \rp $ and $P\lp \sigma^2 \cond  y^{cent,c}_{ik}, \bm{X}_{ik}, A_{ik}, V_k, b_k, \tau^2, \alpha_k, \{\mathcal{W}_h, \mathcal{M}_h\} \rp $  are presented in Web Section S1 of  \cite{hu2022flexible}.  
The posterior distribution of the random intercept $b_k$ is 
 $$\lsq b_k \cond y^{cent,c}_{ik}, \bm{X}_{ik}, A_{ik}, V_k, \tau^2, \alpha_k, \sigma^2, \{\mathcal{W}_h, \mathcal{M}_h\}\rsq \sim N\lp \dfrac{\tau^2 \alpha_k \sum_{i=1}^{n_k}\lp y^{cent,c}_{ik}- \hat{f}(\bm{X}_{ik}, A_{ik})\rp}{n_k\tau^2\alpha_k+\sigma^2},  \dfrac{\sigma^2 \tau^2 \alpha_k}{n_k \tau^2\alpha_k+\sigma^2} \rp.$$
 The posterior of $\alpha_k$,  used for parameter expansion, is 
 $$\lsq \alpha_k \cond y^{cent,c}_{ik}, \bm{X}_{ik}, A_{ik}, V_k, \tau^2, b_k, \sigma^2, \{\mathcal{W}_h, \mathcal{M}_h\}\rsq \sim IG \lp 1, 1+ \frac{\sum_{k=1}^K b_k^2}{2\tau^2}\rp.$$
 The posterior of $\tau^2$ is
 $$\lsq \tau^2 \cond y^{cent,c}_{ik}, \bm{X}_{ik}, A_{ik}, V_k, b_k,\alpha_k, \sigma^2, \{\mathcal{W}_h, \mathcal{M}_h\} \rsq \sim IG \lp \dfrac{K}{2}+1, \dfrac{\sum_{k=1}^K b_k^2+2\alpha_k}{2\alpha_k}\rp.$$ Complete derivation of the posterior distributions are also provided in Web Section S1 of \cite{hu2022flexible}. 
 A \emph{single iteration} of the riAFT-BART sampling algorithm proceeds through the following steps: 
\begin{enumerate}
\item  Update $b_k$, $\tau^2$ and $\alpha_k$ from their respective posterior distributions. \item Using $\log y^{cent,c}_{ik} - b_k$ as the responses and $\{A_{ik}, \bm{X}_{ik}\}$ as the covariates, update BART sum-of-trees  model via  parameters $\mu_{lh}$ and $\sigma^2$, using the Bayesian backfitting approach of  \cite{chipman2010bart}.
Directly update $f(A_{ik}, \bm{X}_{ik})$ using the updated BART model,  for $i = 1, \ldots, n_k, k=1, \ldots, K$. 
\item For each $\{i,k\} \in \{i =1, \ldots, n_k, k =1, \ldots, K\}$, update $z_{ik}$ by sampling 
$$\log z_{ik} \sim \text{Truncated-Normal} \lp f(A_{ik}, \bm{X}_{ik})+b_k, \sigma^2; \log y^{cent}_{ik} \rp.$$
\end{enumerate}

Because we use the centered responses $\log(y^{cent}_{ik}) = \log(y_{ik}) - \hat{\mu}_{AFT}$ in posterior computation, we add $\hat{\mu}_{AFT}$ back to the posterior draws of $f(A_{ik}, \bm{X}_{ik})$ in the final output.

\section{Amputation details for simulation in Section 4.2}

We used the multivariate amputation procedure to generate  incomplete survival datasets with an overall missingness proportion of 40\% in covariates.  For brevity of notation, we suppress subscripts $i$ and $k$. After generating the fully observed data, we amputated predictors  $X_5, X_6, X_7, X_8$  under the missing at random mechanism using the multivariate amputation approach. We first randomly divided the full data into 8 subsamples, which were the following percentages of the whole data: 0.30,  0.09,  0.09, 0.08,  0.08, 0.16,  0.10, and 0.10. The weighted sum scores (WSS) relate the missingness on amputated variables to the values of other variables as follows: 

\begin{enumerate}
    \item[(1)] $wss_{x_5,i}=x_3+x_4+x_3 x_4$
    \item[(2)] $wss_{x_6,i}=x_3+x_4+x_5+x_5^2+x_3 x_4$
    \item[(3)] $wss_{x_7,i}=x_4+x_5+x_6+x_6^2+x_4 x_5$
    \item[(4)] $wss_{x_8,i}=x_5+x_6+x_7+x_6x_7$
    \item[(5)] $wss_{x_5, x_6,i}=x_3+x_4$
     \item[(6)] $wss_{x_6, x_7,i}=x_5$
     \item[(7)] $wss_{x_7, x_8,i}=x_4 + x_5 + 0.5x_4 x_5$
    \item[(8)] $wss_{x_6, x_8,i}=x_3+x_4+x_3 x_4$
\end{enumerate}
The weighted sum score gives a nonzero weight to the variables (the right hand side of the WSS equations) and their nonlinear forms and interactions therein, on which the probabilities to be missing for amputated variables depend.  The predictor variables $X_5, X_6, X_7, X_8$ were respectively amputated in subsample (1), subsample (2), subsample (3), and subsample (4). We further created the joint missingness in $(X_5, X_6)$ in subsample (5), $(X_6, X_7)$ in subsample (6), $(X_7, X_8)$ in subsample (7) and $(X_6, X_8)$ in subsample (8). Finally, we applied the logistic distribution function to the weighted sum scores to create the missing indicators and amputate data. A right-tailed type of missingness was used for subsamples (1)--(5) and a both-tailed type of missingness was used for subsamples (6)--(8). The amputated eight subsamples were combined to form a whole dataset. This procedure created 40\%  overall missingness in the covariates. The missingness proportion is 15\% in $X_5$, 17\% in $X_6$, 14\% in $X_7$ and 11\% in $X_8$.

\section{Supplementary figures and tables}

\begin{figure}[H]
\includegraphics[width = 1\textwidth]{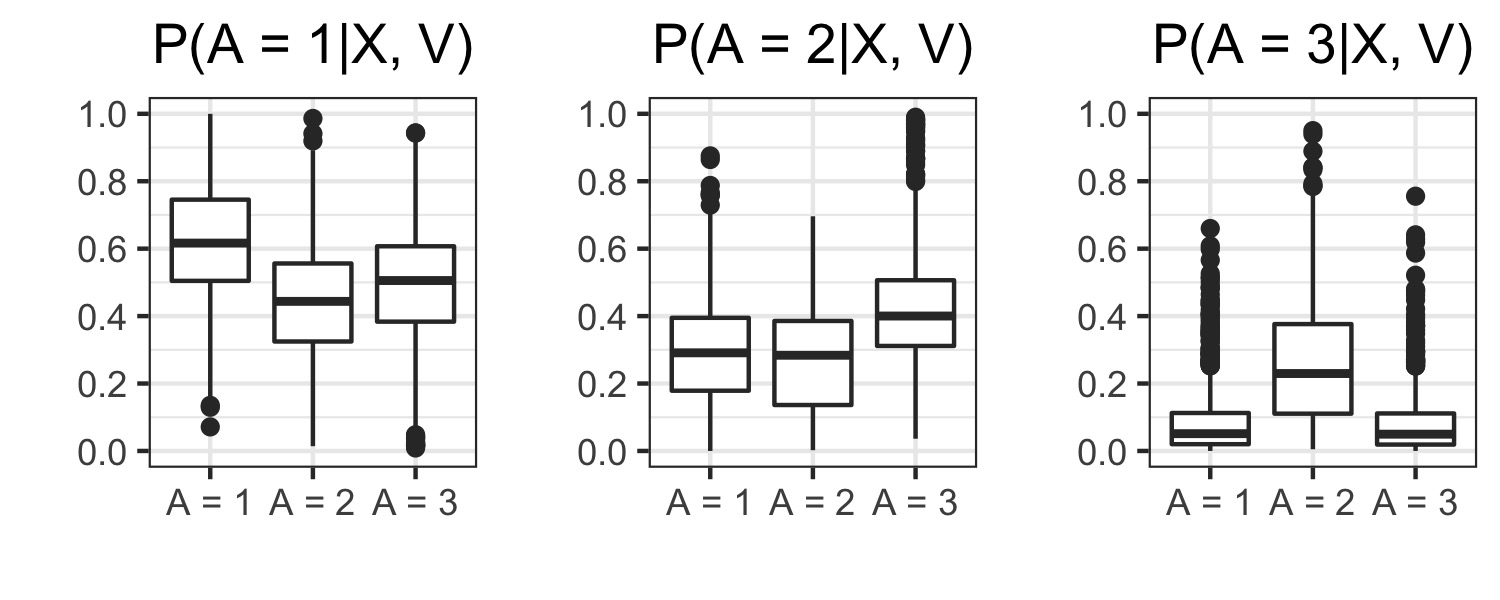}
\caption{Overlap assessment for data simulated under moderate covariate overlap. Each panel presents boxplots by treatment group of the true generalized propensity scores for one of three treatments, and for every unit in the sample. The left panel presents treatment 1, the middle panel presents treatment 2, and the right panel presents treatment 3.}
\end{figure}

\begin{figure}[H]
\includegraphics[width =1\textwidth]{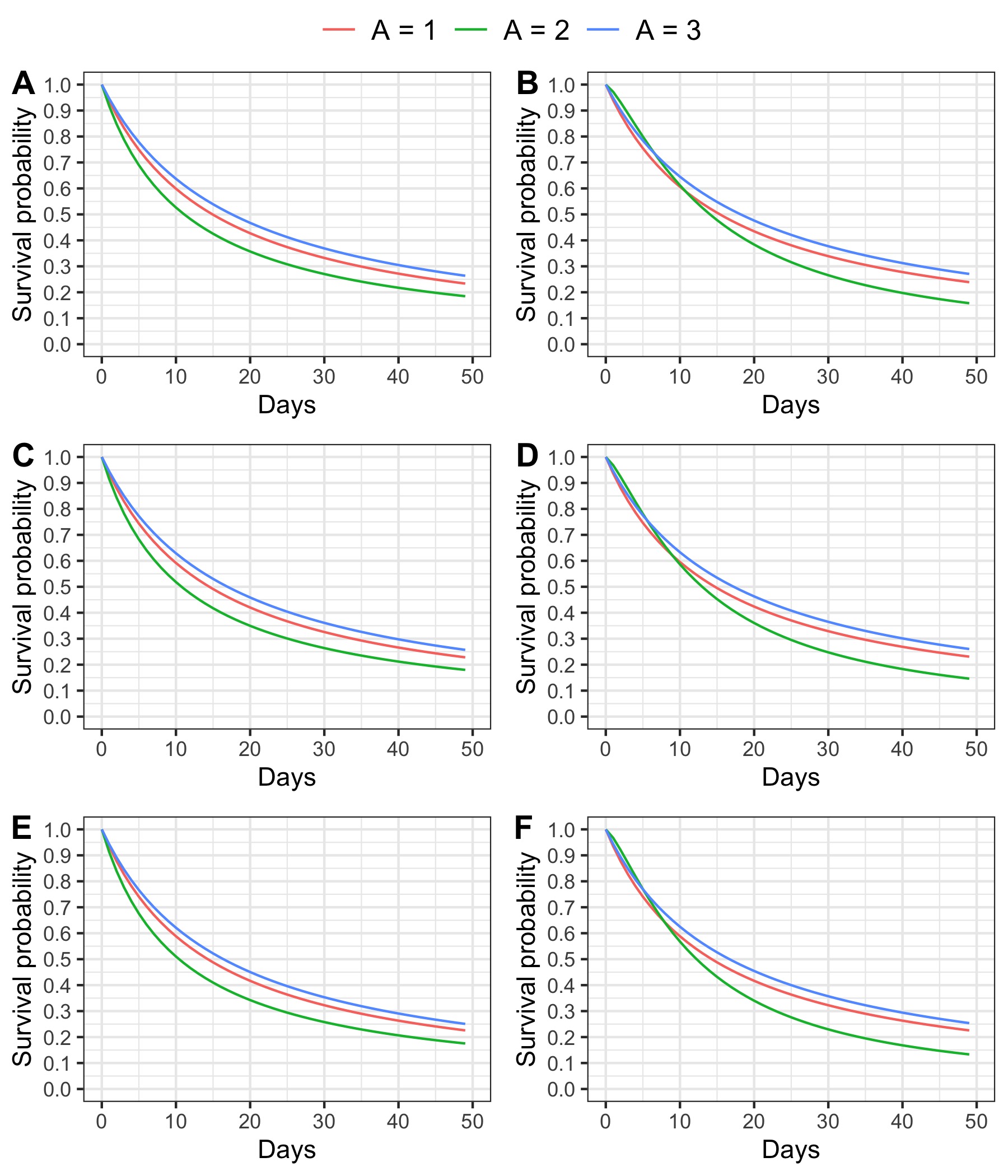}
\caption{Kaplan-Meier survival curves for three treatment groups generated in our simulation study in Section 4.1. Panels A-E respectively represent scenarios corresponding to PH \& HS(a), nPH \&  HS(a), PH \& HS(b), nPH \& HS(b), PH \& HS(c), nPH \& HS(c). PH=proportional hazards; nPH=nonproportional hazards; HS=heterogeneity setting.}
\label{fig:true-suvival}
\end{figure}

\begin{figure}[H]
\includegraphics[width = 1\textwidth]{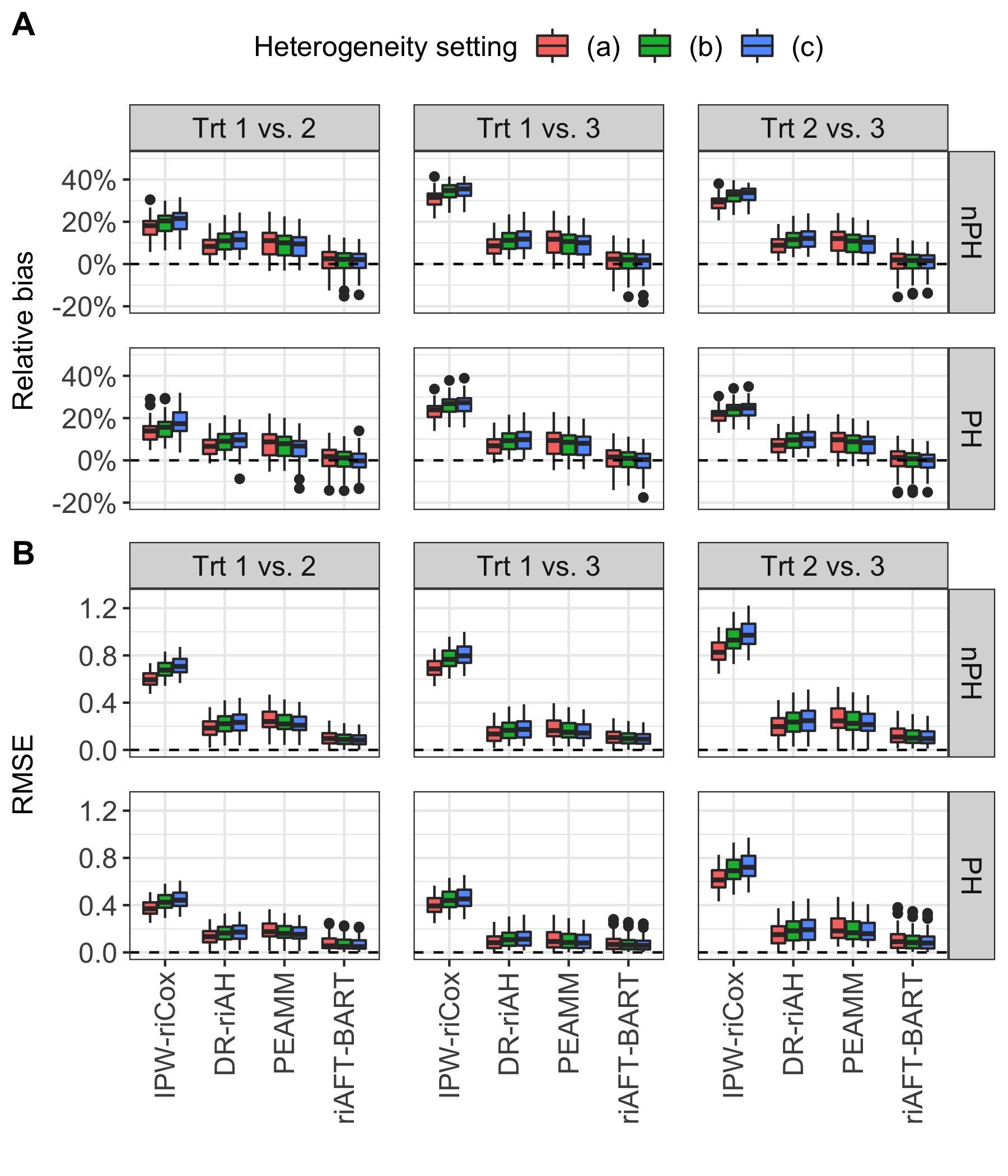}
\caption{Relative biases (Panel A) and root-mean-squared-errors (RMSE) (Panel B) among 40 generalized propensity score subgroups under 6 data configurations: (heterogeneity settings a, b, c) × (proportional hazards (PH) and nonproportional hazards (nPH)) for each of four methods, IPW-riCox, DR-riAH, PEAMM and riAFT-BART. Three pairwise treatment effects were estimated by averaging the individual survival treatment effect (based on 3-week restricted mean survival time) across individuals in each subgroup.  Each boxplot visualizes the distribution of relative biases or the distribution of RMSE for 40 subgroups, each averaged across 250 simulation runs.}
\end{figure}

\begin{figure}[H]
    \centering
    \includegraphics[width=1\textwidth]{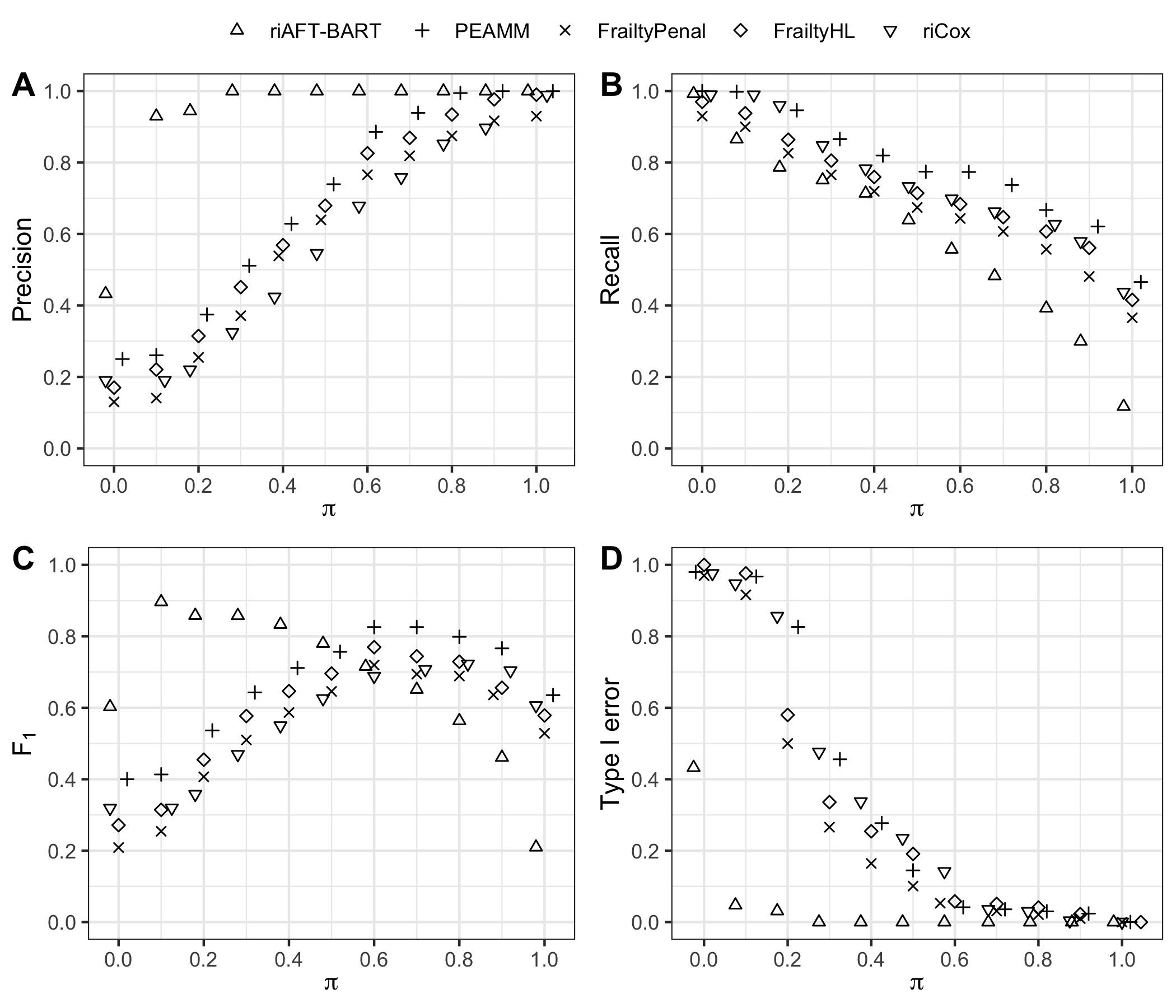}
    \caption{The precision, recall, $F_1$ score and Type I error, for each of four methods: riAFT-BART, PEAMM, riCox, FrailtyHL and FrailtyPenal with data generated under proportional hazards, based on 250 data replications. Imputation was performed on 100 bootstrap samples of each replication dataset, using imputation method $\code{mice}$. There are $K=10$ clusters, each with a size of 200; the total sample size is 2000. The overall proportion of missingness is 40\%. } 
    \label{fig:pi_metrics_PH}
\end{figure}

\begin{figure}[H]
    \centering
    \includegraphics[width=1\textwidth]{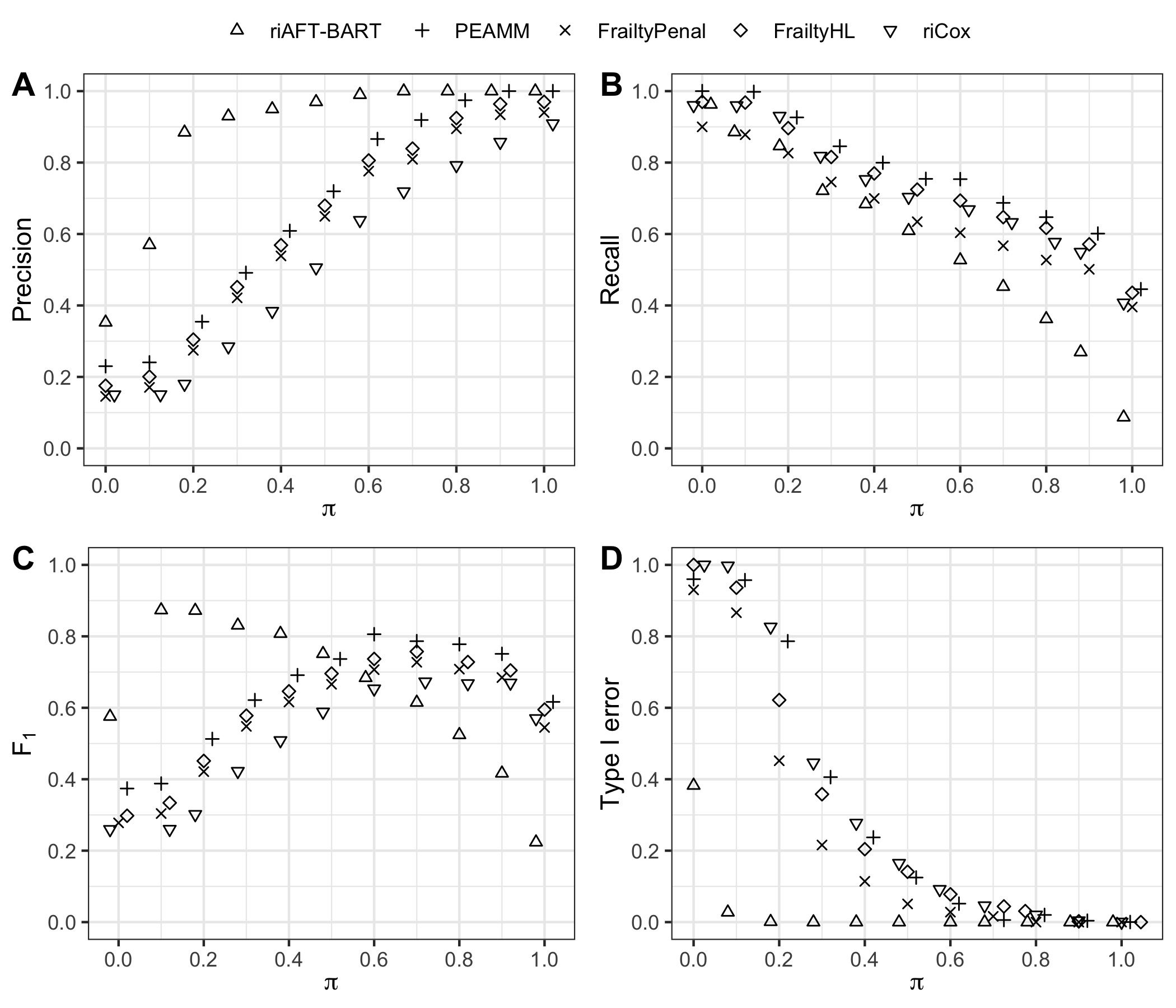}
    \caption{The precision, recall, $F_1$ score and Type I error, for each of four methods: riAFT-BART, PEAMM, riCox, FrailtyHL and FrailtyPenal with data generated under non-proportional hazards, based on 250 data replications. Imputation was performed on 100 bootstrap samples of each replication dataset, using imputation method $\code{mice}$. There are $K=10$ clusters, each with a size of 200; the total sample size is 2000.The overall proportion of missingness is 40\%.  } 
    \label{fig:pi_metrics_nPH}
\end{figure}

\begin{figure}[H]
    \centering
    \includegraphics[width=1\textwidth]{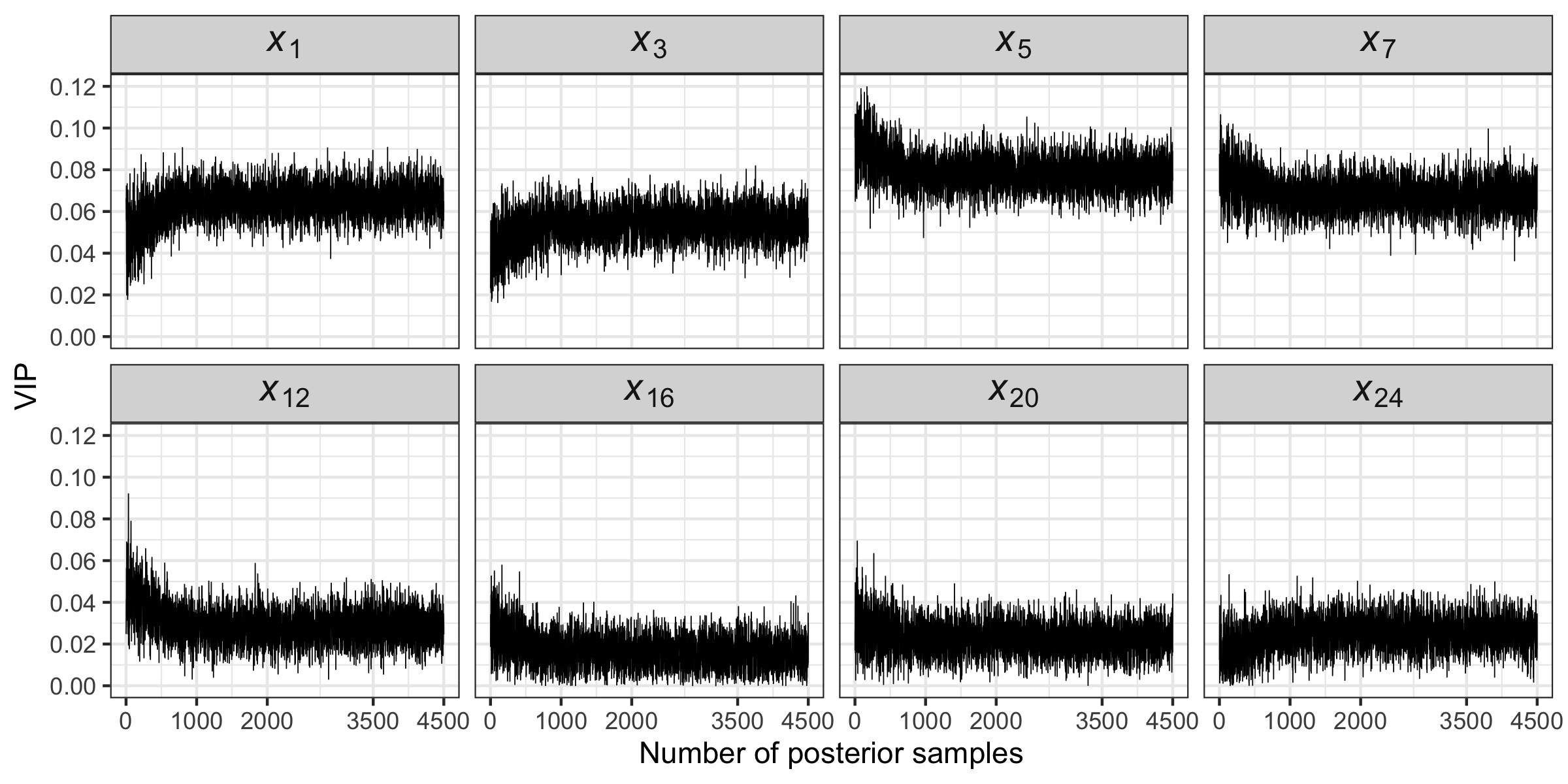}
    \caption{Assessing convergence of the chain by plotting 4500 posterior draws of variable selection proportions (VIP)  for 4 useful predictors $X_{ik1}, X_{ik3}, X_{ik5}, X_{ik7}$ and 4 noise predictors $X_{ik12}, X_{ik16}, X_{ik20}, X_{ik24}$. The first 1000 posterior draws are discarded as burn-in.}
\end{figure}

\begin{figure}[H]
\includegraphics[width = 1\textwidth]{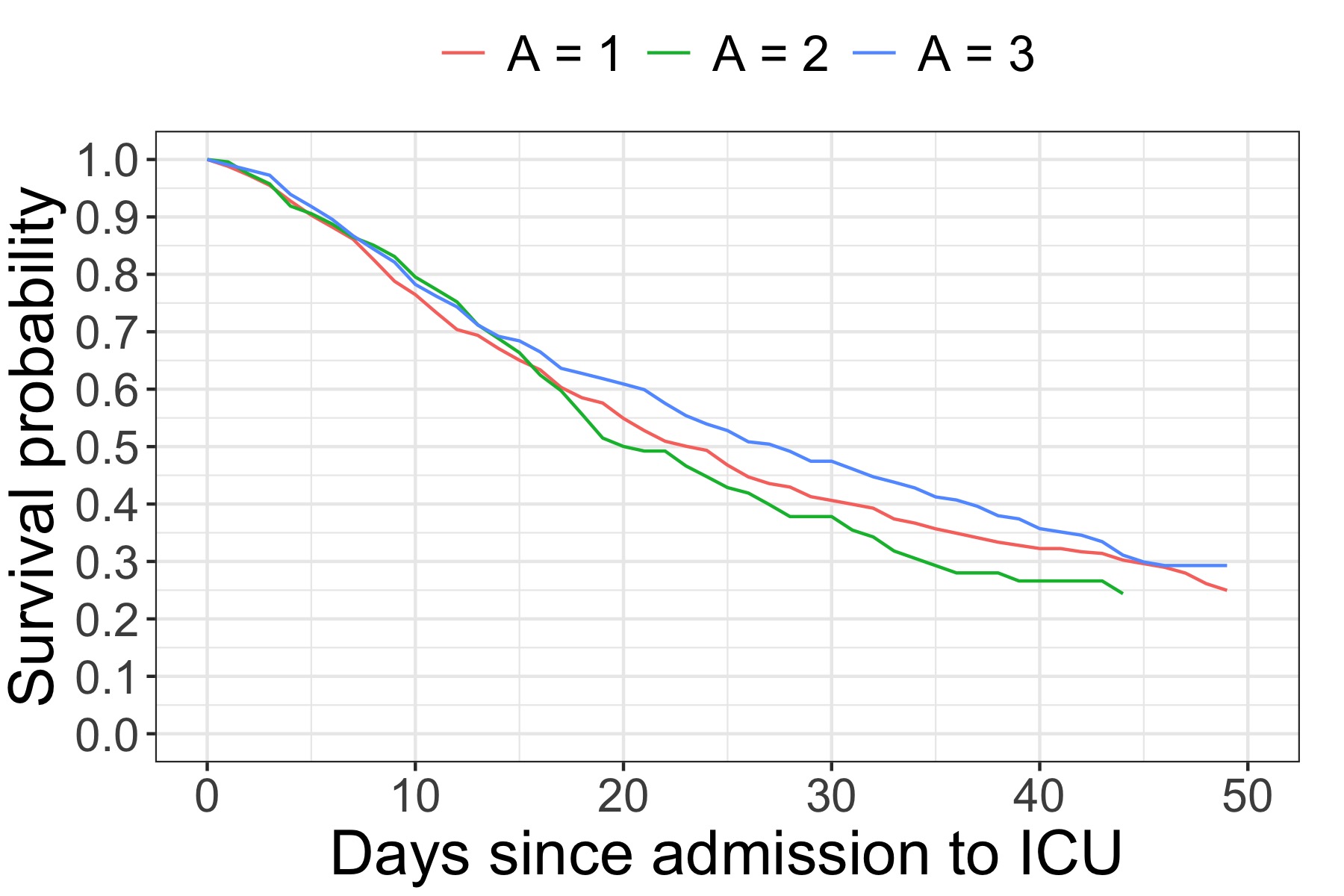}
\caption{The Kaplan-Meier survival curves for three treatment groups. Three treatment options are $A = 1$: Dexamethasone, $A = 2$: Remdesivir and $A = 3$: Dexamethasone + Remdesivir.}
\label{fig:KM-postICU}
\end{figure}

\begin{figure}[H]
\includegraphics[width = 1\textwidth]{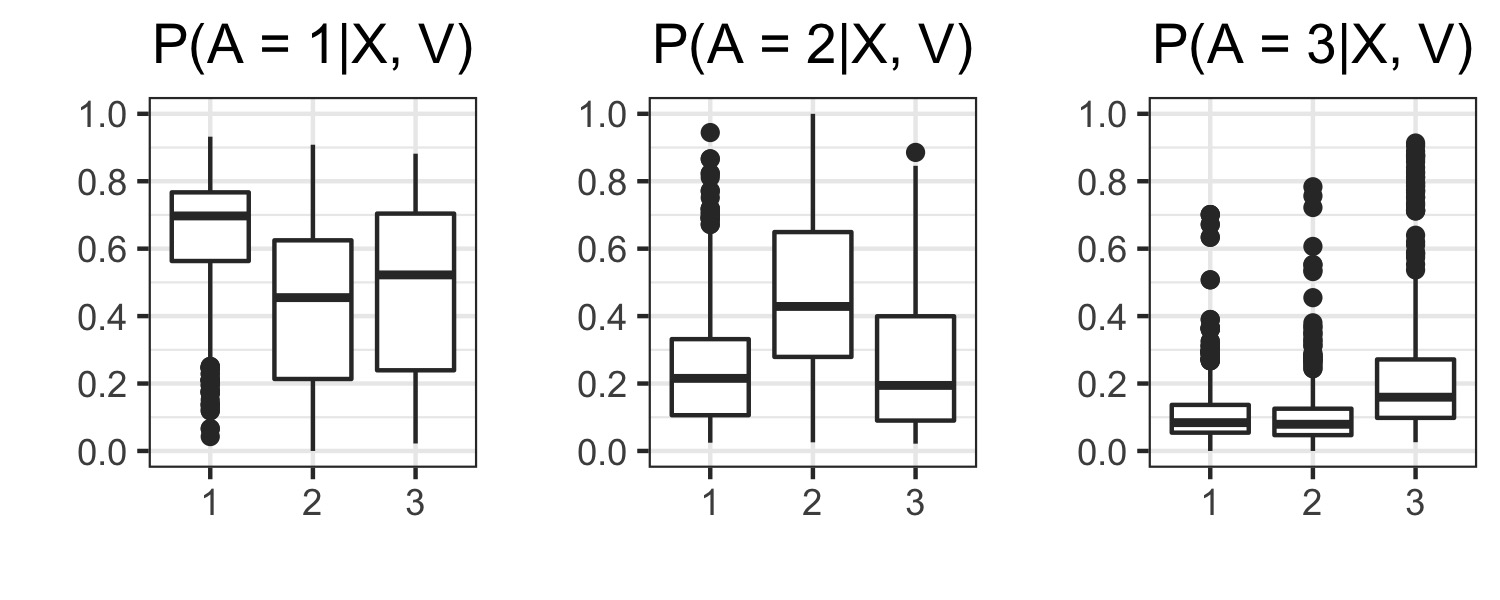}
\caption{Overlap assessment for three treatment groups in the COVID-19 dataset. Each panel presents boxplots by treatment group of the generalized propensity scores, estimated by Super Learner, for one of three treatments, and for every individual in the sample. The left panel presents treatment 1 = Dexamethasone, the middle panel presents treatment 2 = Remdesivir, and the right panel presents treatment 3 = Dexamethasone + Remdesivir.}
\label{fig:overlap-postICU}
\end{figure}

\begin{figure}[H]
\centering 
\includegraphics[width = 1\textwidth]{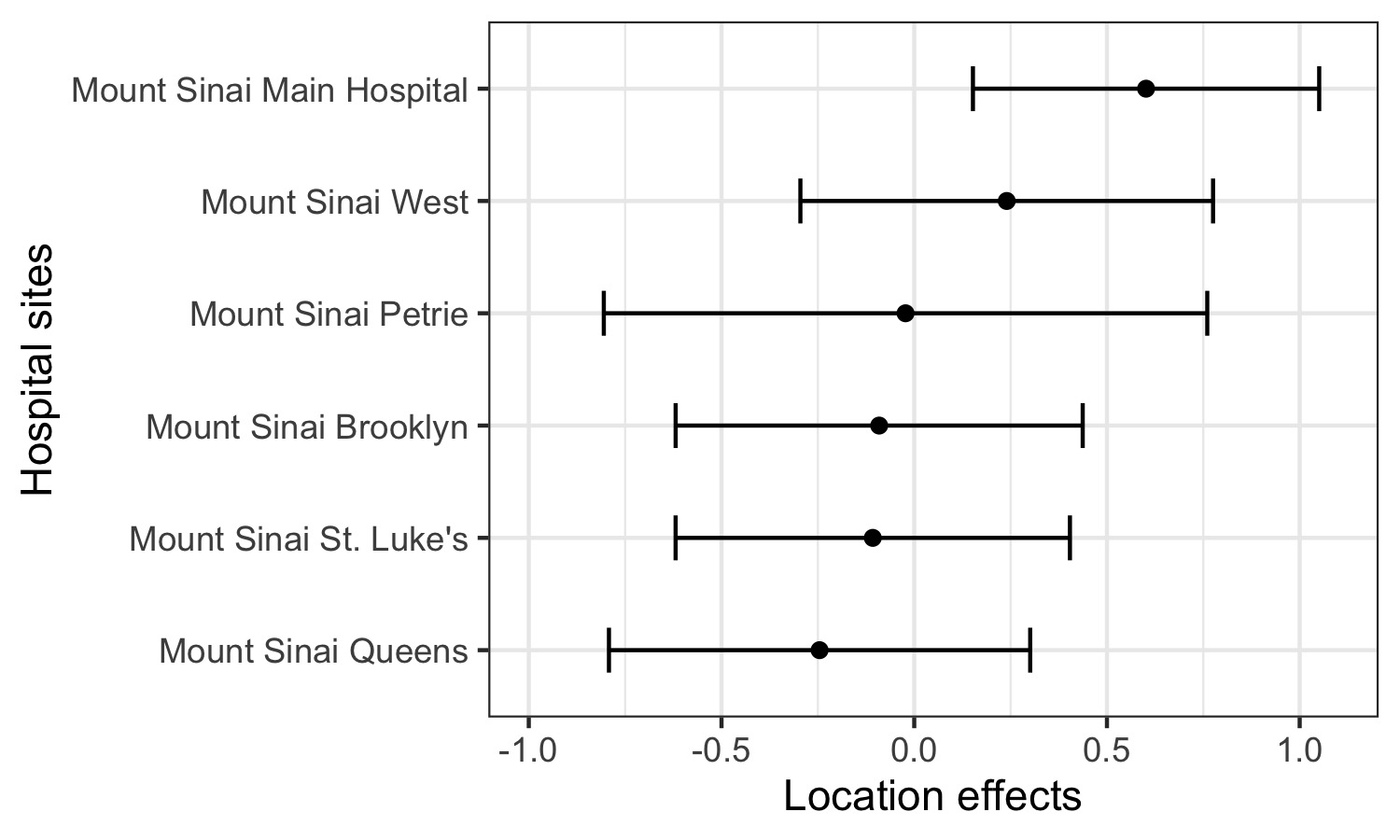}
\caption{Examining the effect of hospital sites on patient survival in terms of the log survival days represented by the posterior mean and credible intervals of the random intercept $b_k$, $k=1, \ldots, 6$, for COVID-19 case study.}
\label{fig:covid_location}
\end{figure}

\begin{table}[H]
\centering
\caption{Subclasses of the true generalized propensity scores (GPS) described in our simulation study in Section 4.1. The number of individuals falling in each subclass was calculated based on one data replication. }
\begin{tabular}{cccc} 
\toprule
  Subclass ID & GPS for treatment group 1 & GPS for treatment group 2 & \# of individuals \\
 \midrule
  1 & $(0, 0.1]$ & $(0, 1]$ & 60\\
  2 & $(0, 0.2]$ & $(0, 0.25]$ & 52 \\
  3 & $(0, 0.2]$ & $(0.25, 0.5]$ & 48\\
  4 & $(0, 0.2]$ & $(0.5, 0.75]$ & 52\\  
  5 & $(0, 0.2]$ & $(0.75, 1]$ & 48\\  
  6 & $(0.2, 0.4]$ & $(0, 0.2]$  & 44\\  
  7 & $(0.2, 0.4]$ & $(0.2, 0.4]$ & 45\\  
  8 & $(0.2, 0.4]$ & $(0.4, 0.6]$ & 56\\  
  9 & $(0.2, 0.4]$ & $(0.6, 0.8]$ & 42\\  
  10 & $(0.2, 0.4]$ & $(0.8, 1]$ & 41\\  
  11 & $(0.4, 0.5]$ & $(0, 0.1]$ & 42\\  
  12 & $(0.4, 0.5]$ & $(0.1, 0.3]$ & 44\\  
  13 & $(0.4, 0.5]$ & $(0.3, 0.4]$ & 46\\  
  14 & $(0.4, 0.5]$ & $(0.4, 0.5]$ & 54\\  
  15 & $(0.4, 0.5]$ & $(0.5, 0.6]$ & 56\\  
  16 & $(0.4, 0.5]$ & $(0.6, 0.7]$ & 53\\  
  17 & $(0.4, 0.5]$ & $(0.7, 1]$ & 50\\  
  18 & $(0.5, 0.6]$ & $(0, 0.2]$ & 42\\  
  19 & $(0.5, 0.6]$ & $(0.2, 0.3]$ & 44\\  
  20 & $(0.5, 0.6]$ & $(0.3, 0.4]$ & 48\\  
  21 & $(0.5, 0.6]$ & $(0.4, 0.5]$ & 56\\  
  22 & $(0.5, 0.6]$ & $(0.5, 0.6]$ & 57\\  
  23 & $(0.5, 0.6]$ & $(0.6, 0.7]$ & 46\\  
  24 & $(0.5, 0.6]$ & $(0.7, 0.8]$ & 44\\  
  25 & $(0.5, 0.6]$ & $(0.8, 1]$ & 45\\  
  26 & $(0.6, 0.7]$ & $(0, 0.3]$ & 48\\
  27 & $(0.6, 0.7]$ & $(0.3, 0.5]$& 45 \\
  28 & $(0.6, 0.7]$ & $(0.5, 0.6]$ & 55\\
  29 & $(0.6, 0.7]$ & $(0.6, 0.7]$ & 52\\
  30 & $(0.6, 0.7]$ & $(0.7, 1]$ & 50\\
  31 & $(0.7, 0.8]$ & $(0, 0.3]$ & 45\\
  32 & $(0.7, 0.8]$ & $(0.3, 0.5]$ & 54 \\
  33 & $(0.7, 0.8]$ & $(0.5, 0.7]$ & 52\\
  34 & $(0.7, 0.8]$ & $(0.7, 1]$ & 46\\
  35 & $(0.8, 0.9]$ & $(0, 0.4]$ & 57\\
  36 & $(0.8, 0.9]$ & $(0.4, 0.6]$ & 48\\
  37 & $(0.8, 0.9]$ & $(0.6, 1]$ & 58\\
  38 & $(0.9, 1]$ & $(0, 0.4]$ & 63\\
  39 & $(0.9, 1]$ & $(0.4, 0.6]$& 52 \\
  40 & $(0.9, 1]$ & $(0.6, 1]$ & 60\\
\bottomrule
\end{tabular}
\end{table}

\begin{table}[H]
\centering
\caption{Mean and (standard deviation) of precision in the estimation of heterogeneity effects (PEHE) across 250 data replications for each of the 4 methods based on 3-week restricted mean survival time under six configurations: (proportional hazards (PH) vs. nonproporitonal hazards (nPH)) $\times$ (heterogeneity setting (a) vs. (b) vs. (c)).}
\begin{tabular}{clccccccc}
\toprule
& &  \multicolumn{3}{c}{PH} && \multicolumn{3}{c}{nPH}  \\
\cmidrule{3-5}  \cmidrule{7-9}
 & Methods &  HS(a) & HS(b) & HS(c) & & HS(a) & HS(b) & HS(c) \\
\midrule
\multirow{4}{*}{Trt 1 vs. 2}  & IPW-riCox &2.32 (.22)	&	2.65 (.25) &  2.77 (.30) & & 3.12 (.27)	&	3.45 (.30) &  3.55 (.30) \\
& DR-riAH & 0.67 (.05)	& 0.75 (.05) & 0.79 (.05) && 0.82 (.05)	&	0.90 (.06) & 0.94 (.06)\\
&PEAMM & 0.89 (.06)	& 0.81 (.05)	 & 0.74 (.05) && 1.08 (.07)	&   1.00 (.06)	 &  0.91 (.06)\\
&riAFT-BART & 	0.28 (.04) & 0.24 (.03) & 0.19 (.03)   && 0.37 (.04)  & 0.32 (.03)& 0.27 (.03)\\
\midrule
\multirow{4}{*}{Trt 1 vs. 3}  & IPW-riCox &2.43 (.23)	&	2.75 (.23) &  2.84 (.31) & & 3.23 (.28)	&	3.56 (.31) &  3.64 (.31) \\
& DR-riAH & 0.73 (.05)	& 0.76 (.05) & 0.83 (.05) && 0.85 (.05)	&	0.94 (.06) & 0.99 (.06)\\
&PEAMM & 0.93 (.06)	& 0.85 (.05)	 & 0.78 (.05) && 1.12 (.07)	&   1.02 (.06)	 &  0.93 (.06)\\
&riAFT-BART & 	0.32 (.04) & 0.27 (.03) & 0.23 (.03)   && 0.40 (.04)  & 0.35 (.03)& 0.30 (.03)\\
\midrule
\multirow{4}{*}{Trt 2 vs. 3}  & IPW-riCox & 2.50 (.23)	&	2.83 (.26) &  2.94 (.31) & & 3.30 (.28)	&	3.64 (.31) &  3.73 (.31) \\
& DR-riAH & 0.77 (.05)	& 0.84 (.05) & 0.87 (.05) && 0.89 (.05)	&	0.98 (.06) & 1.01 (.06)\\
&PEAMM & 0.97 (.06)	& 0.89 (.05)	 & 0.82 (.05) && 1.16 (.07)	&   1.06 (.06)	 &  0.97 (.06)\\
&riAFT-BART & 	0.35 (.04) & 0.30 (.03) & 0.27 (.03)   && 0.43 (.04)  & 0.39 (.03)& 0.35 (.03)\\
\bottomrule
\end{tabular}
\end{table}

\begin{table}[H]
 \caption{The definition of patient oxygen levels based on the use of ventilator.} 
 \def\arraystretch{1.25} 
\begin{tabular}{p{3.5cm} p{12.5cm}}
\toprule
Patient oxygen level   & Ventilator status \\
 \midrule
0 & Room air\\
1 & Cannula\\
2 & Mask, Blow-by, Face tent, Oxyhood, Non-rebreather, RAM cannula\\
3 & Continuous positive airway pressure machine, High flow nasal cannula, Hudson prongs\\
4 & Bilevel positive airway pressure machine, Tracheostomy mask\\
5 & Tracheotomy, Transtracheal oxygen therapy, Ventilator, Endotracheal tube, T-shaped tubing connected to an endotracheal tube, Nasal synchronized intermittent mandatory ventilation\\
 \bottomrule
\end{tabular}\label{tab:def_pol}
\end{table}

\begin{table}[H]
\centering
\footnotesize
\def\arraystretch{0.7} 
\caption{Baseline variables of the COVID-19 data. Summary statistics are represented as mean and standard deviation (SD) for continuous variables and N (\%) for discrete variables.}
\begin{tabular}{lcccc}
\toprule 
& Overall  & Dexamethasone  & Remdesivir  & Dexamethasone + Remdesivir \\ 
Characteristics & $N = 1955$ & $N = 1097$ & $N = 620$ & $N = 238$ \\
\hline
Age (years)  & 65.13 (14.95) & 65.11 (13.64) & 63.48 (16.68) & 69.49 (15.15) \\  
Gender  &\\
\;\;  Male &     1223 (62.6) &    688 (62.7) &    391 (63.1)  &    144 (60.5)  \\ 
\;\;  Female &  732 (37.4) &  409  (37.3) & 229 (36.9) &   94 (39.5)  \\  
Race  &\\
\;\; White &   562 (28.7)  &   336 (30.6)  &  151 (24.4)   &    75 (31.5)   \\ 
\;\; Black &    428 (21.9) &   195 (17.8)  &   190 (30.6)  &    43 (18.1)  \\ 
\;\;   Asian  &      110 (5.6)   &    59 (5.4)  &   29 (4.7)   &    22 (9.2)  \\ 
\;\;   Others  &     855 (43.7)   &   507 (46.2)  &   250 (40.3)  &    98 (41.2)  \\ 
Ethnicity  &\\
\;\;  Hispanic &  504 (25.8)  &  305 (25.8) &  146 (23.5) & 53 (22.3)  \\ 
\;\; Non-Hispanic &   1451 (74.2)  &   792 (72.2)  &   474 (76.5)  &   185 (77.7)  \\ 
Asthma  &\\
\;\;  Yes &     205 (10.5)   &   137 (12.5)  &   48 (7.7)   &    20 (8.4)     \\ 
\;\;  No &  1750 (89.5) & 960 (87.5) & 572 (92.3) & 218 (91.6)\\ 
COPD  &\\
\;\;  Yes &  142 (7.3)   &    96 (8.8)  &   31 (5.0)   &   15 (6.3)    \\ 
\;\;  No & 1813 (92.7)  &  1001 (91.2) & 589 (95.0) & 223 ( 93.7)\\ 
Hypertension  &\\
\;\;  Yes &  1025 (52.4)  &   572 (52.1)  &  323 (52.1)   &   130 (54.6)     \\ 
\;\;  No &  930 (47.6) & 525 (47.9) &  297 (47.9) &  108 (45.4)\\ 
Cancer  &\\
\;\;  Yes & 187 (9.6)   &    93 (8.5)  &     65 (10.5) &    29 (12.2)   \\ 
\;\;  No & 1768 (90.4)  & 1004 (91.5) & 555 (89.5) & 209 (87.8)\\ 
Coronary artery disease  &\\
\;\;  Yes &    433 (22.1)   &   237 (21.6)  &  142 (22.9)   &     54 (22.7)   \\ 
\;\;  No & 1522 (77.9) & 860 (78.4) & 478 (77.1) & 184 (77.6)\\ 
Diabetes  &\\
\;\;  Yes &  536 (27.4)  &   300 (27.3)  &   170 (27.4)  &    66 (27.7) \\ 
\;\;  No &   1419 (72.6)  & 797 (72.7) & 450 (72.6)  & 172 (72.3)\\ 
Smoking status  &\\
\;\;  Current &      85 (4.3)  &   43 (3.9)   &    36 (5.8)  &    6 (2.5)  \\
\;\;  Former &     455 (23.3)   &  236 (21.5)   &   146 (23.5)  &    73 (30.7)  \\ 
\;\;  Never &      982 (50.2)   &   553 (50.4)  &   306 (49.4)  &  123 (51.7)  \\ 
\;\;  Unknown &     433 (22.1)  &  265 (24.2)   &   132 (21.3)  &    36 (15.1)   \\ 

Patient oxygen level  &\\
\;\;  0 &        114 (5.8)   &      11 (1.0)    &      94 (15.2)   &        9 (3.8)   \\
\;\;  1 &      149 (7.6)   &     38 (3.5)    &     92 (14.8)   &       19 (8.0)   \\ 
\;\;  2 &        110 (5.6)   &      36 (3.3)   &     59 (9.5)   &     15 (6.3)  \\ 
\;\;  3 &       396 (20.3)   &   269 (24.5)    &      45 (7.3)  &     82 (34.5)   \\ 
\;\;  4 &    273 (14.0)   &   199 (18.1)  &     36 (5.8)  &       38 (16.0)   \\ 
\;\;  5 &      913 (46.7)  &   544 (49.6)   &    294 (47.4)  &       75 (31.5)    \\ 
D dimer (ng/mL)  &  4.34 (4.70) &  4.35 (4.77) &  4.47 (4.69) &  3.93 (4.43) \\
Creatinine (mg/dL)&   1.77 (1.93) &  1.67 (1.73) &   1.92 (2.12) &  1.83 (2.23) \\
Systolic BP (mgHg)&144 (23) &140 (22) & 147 (25) & 143 (21)\\
Diastolic BP (mgHg)&83 (16) &81 (15) & 87 (15) & 80 (14)\\
Temperature ($^\circ C$)& 37.8 (0.6) & 37.4 (0.8) & 37.9 (0.7) & 37.5 (0.9)\\
White Blood Cell (K/ug) & 7.6 (2.2)& 7.5 (2.3)& 7.9(2.6) & 7.4(2.7)\\
C-reactive protein (mg/L) & 113 (24) & 110 (22) & 116 (24) & 112 (28) \\
SOFA score & 2.56 (2.84) & 2.36 (2.34) & 2.86 (2.24)  & 2.58 (2.74) \\
Lactate dehydrogenase (U/L) & 412 (34) & 436 (37)  & 406 (26)  & 416 (27) \\
Heart rate (beats/min) & 102 (6) & 100 (5) & 105 (7) & 103 (4)\\
Ferritin (ng/ml) & 708 (103) & 803 (109) & 610 (83) & 712 (97)\\
Glasgow coma scale & 11.3 (5.2) & 10.6 (4.6)  & 11.8 (5.2) & 11.2 (5.1)\\
Body Mass Index (Kg/$m^2$) & 27.4 (1.2) & 27.7 (1.3) & 27.0 (1.7) & 26.8 (1.5)\\
Oxygen saturation (\%) & 94.7 (5.6) & 94.3 (5.0) & 94.0 (5.0) & 95.0 (5.0)\\
Fraction of inspired oxygen & 0.35 (0.03) & 0.31 (0.03) & 0.37 (0.04) & 0.42 (0.04)\\
Partial pressure of oxygen (mmHg) & 118 (29) & 129 (34)  & 110 (30)  & 120 (31)\\  
Hospital site  &\\
\;\;     Mount Sinai Brooklyn &      272 (13.9)  &  103 (9.4)   &   119 (19.2)  &    50 (21.0)  \\
\;\;     Mount Sinai Petrie &        58 (3.0)    &     7 (0.6)  &   12 (1.9)   &    39 (16.4)  \\ 
\;\;   Mount Sinai Queens &       245 (12.5)   &   166 (15.1)  &    57 (9.2)  &   22 (9.2)  \\ 
\;\;     Mount Sinai St. Luke's &    326 (16.7)  &  157 (14.3)   &   134 (21.6)  &    35 (14.7)    \\ 
  \;\;   Mount Sinai West &    225 (11.5)   &  111 (10.1)   &   92 (14.8)   &   22 (9.2)  \\ 
  \;\;   Mount Sinai Main Hospital &  829 (42.4)  &  553 (50.4)   &   206 (33.2)  &    70 (29.4)  \\ 
 \bottomrule
\end{tabular}
\footnotesize \\\smallskip
Abbreviations:  BP = blood pressure; SOFA = Sequential organ failure assessment
\end{table}

\begin{table}[H]
\centering
\caption{The posterior mean and 95\% credible intervals for three pairwise sample population treatment effects on patient survival for the COVID-19 case study. The effects were based on the difference in the expected log survival days. Three treatment options are $A = 1$: dexamethasone, $A = 2$: remdesivir and $A = 3$: dexamethasone + remdesivir.}
\label{tab:CATE}
\begin{tabular}{ccccc} 
\toprule
  &$CATE_{1,2}$ & $CATE_{1,3}$ & $CATE_{2,3}$ \\
 \midrule
  &  $0.17 (-0.34,0.68)$ & $-0.32 (-0.82,0.18)$ & $-0.49 (-0.95, -0.03)$\\
\bottomrule
\end{tabular}
\end{table}

\section{Data \& Code}
We provide comprehensive instructions for executing our $\R$ scripts, which will enable the reproduction of all tables and figures presented in the manuscript. All $\R$ scripts can be accessed at this GitHub repository \url{https://github.com/liangyuanhu/TEH-VS-riAFTBART}. The specific steps are as follows:

\begin{itemize}
    \item Table 1: Execute `\code{simulation\_TEH.R}'.

\item Table 2:  Execute `\code{sim\_cox\_var\_select.R}' for riCox results,  `\code{sim\_frailtyHL\_var\_select.R}' FrailtyHL results,  `\code{sim\_frailtypack\_var\_select.R}' for FrailtyPenal results, `\code{sim\_peamm\_var\_select.R}' for PEAMM results,  `\code{sim\_riaftbart\_var\_select.R}' for riAFT-BART results.

\item Table 3: Execute `\code{case\_study\_comparison.R}'. 

\item Figure 1: Execute `\code{simulation\_TEH.R}', followed by `\code{figure\_1.R}'.  

\item Figure 2: Execute `\code{sim\_cox\_var\_select.R}', `\code{sim\_frailtyHL\_var\_select.R}', \\`\code{sim\_frailtypack\_var\_select.R}', `\code{sim\_peamm\_var\_select.R}', `\code{sim\_riaftbart\_var\_select.R}', followed by  `\code{figure\_2.R}'.

\item Figure 3: Execute `\code{sim\_cox\_var\_select.R}', `\code{sim\_frailtyHL\_var\_select.R}', \\`\code{sim\_frailtypack\_var\_select.R}', `\code{sim\_peamm\_var\_select.R}', `\code{sim\_riaftbart\_var\_select.R}', followed by `\code{figure\_3.R}'.

\item Figure 4: Execute `\code{case\_study\_comparison.R}', followed by `\code{figure\_4.R}'.

\item Figure 5: Execute `\code{case\_study\_riAFTBARTR.R}', followed by `\code{figure\_5.R}'.
\end{itemize}
\newpage
\bibliography{references}